\documentclass[notitlepage, amsfonts, amssymb, amsmath, showkeys, nofootinbib,longbibliography,aps,prd]{revtex4-1}
\usepackage[english]{babel}
\usepackage[utf8]{inputenc}
\usepackage{tabularx}
\usepackage{graphicx}
\usepackage{adjustbox}
\usepackage{booktabs}   
\usepackage{ltablex}
\usepackage{enumitem}
\usepackage{fancyhdr}

\usepackage[pdftex, pdftitle={Article}, pdfauthor={Author}]{hyperref} % For hyperlinks in the PDF
\begin{document}
\title{Calorimetric approach to paleo-detection of dark matter}
% SGS requires a running title of <=50 characters

\author{S. Hedges}
\author{P. Huber}\email[]{pahuber@vt.edu}
    \affiliation{Center for Neutrino Physics, Virginia Tech, Blacksburg, VA, USA}

\date{\today} % Leave empty to omit a date

\begin{abstract}
We present the first paleo-detector dark matter sensitivity analysis
based on a calorimetric readout, in which the number of stable lattice
vacancies produced by each nuclear recoil is used as a per-event
observable complementary to the track length. Using full-cascade SRIM
simulations in olivine, we compute the expected sensitivity for a
$100\,\mathrm{g}\cdot 1\,\mathrm{Gyr}$ exposure. We find that a
vacancy-only readout reaches a sensitivity envelope very similar to
that of state-of-the-art track-only analyses. The combination of the
two observables provides an event-by-event proxy for $|dE/dx|$ and
hence for the recoiling nuclear species. Since the neutron-nucleus
cross section is approximately flat in nuclear mass while the
dark-matter--nucleus cross section scales as $A^2$, this
discrimination suppresses the dominant neutron background
by more than an order of magnitude at moderate dark matter masses. The combined-analysis sensitivity reaches
spin-independent dark-matter--nucleon cross sections of order
$10^{-48}\,\mathrm{cm}^2$ at WIMP masses of a few tens of GeV,
comparable to future direct detection experiments. A two-stage readout
combining selective-plane illumination microscopy with scanning
electron microscopy is identified as a path to making a
$100\,\mathrm{g}$-scale analysis plausible.
\end{abstract}

\maketitle

\section{Introduction}
\label{sec:intro}

Persistent structural damage from nuclear interactions in solids has
been used since the 1960s to date minerals~\cite{price1963fossil}. The
fission of uranium impurities in mica, apatite, and olivine produces
tracks of order tens of micrometers in length which, once formed,
remain stable for hundreds of millions of years and can be revealed by
chemical etching and counted under an optical
microscope~\cite{Wagner:1992,Gleadow:2002}. Thus minerals are natural
detectors of nuclear events that have occurred in their volume during
their lifetime. This has led to the idea that they can be used as
detectors of rare events as well, in particular of speculative
hitherto undiscovered new particles. First proposals looked for
magnetic monopoles~\cite{fleischer1969monopoles,price1986supermassive}
to be later followed by concepts for dark matter
detection~\cite{SnowdenIfft:1995}. These methods are in aggregate
referred to as \emph{paleo-detectors}. Paleo-detection has seen a
resurgence in
interest~\cite{Drukier:2018pdt,Baum:2018tfa,Baum:2023cct}; in
particular with the renewed interest in light dark-matter, that was
dismissed prior to the LHC results. The basic idea is simple: a
$100\,\mathrm{g}$ sample of $1\,\mathrm{Gyr}$ old mineral corresponds
to an integrated exposure of $100\,\mathrm{kt}\cdot\mathrm{yr}$, three
orders of magnitude larger than currently planned liquid noble gas
direct-detection experiments~\cite{DARWIN:2016hyl,XLZD:2024}. Apart
from the potentially large exposure, paleo-detection also allows us to
learn about the radiation history of the sample in terms of atmospheric
neutrinos~\cite{Jordan:2020gxx}, solar
neutrinos~\cite{Tapia-Arellano:2021cml}, star
formation~\cite{Baum:2022wfc} and cosmic
rays~\cite{Caccianiga:2024otm,Galelli:2025gss}.

The dominant approach studied in the literature for extracting physics
from this exposure has been to focus on the {\it track length}
produced by each
recoil~\cite{Baum:2018tfa,Baum:2021jak,Baum:2023cct}. In the initial phenomenological estimate, the track length left by a primary
knock-on atom is approximated by integrating the inverse stopping
power $|dE/dx|^{-1}$ from zero to the recoil energy and depends on the
recoiling species through $|dE/dx|$. More recently, fully simulated
events have been used to determine the track length and with it the effect
of fluctuations in track length has been studied as
well~\cite{Fung:2025cub}.

The practical details of sub-micron track detection in gram-quantities
of material have not been fully developed yet, but candidate
technologies include scanning and transmission electron microscopy,
small-angle X-ray scattering, helium-ion beam microscopy, atomic force
microscopy, and others, see Ref.~\cite{Baum:2023cct} for a
review. These approaches yield strong sensitivity at moderate-to-high
recoil energies but lose discriminating power at low recoil energy,
where tracks become shorter than the resolution of the readout method.

The same underlying physics, however, provides an entirely
complementary observable. Each recoil produces not only a track of
finite length but also a population of stable lattice vacancies along
that track with the total number scaling roughly with the deposited
energy. Counting vacancies is therefore a calorimetric measurement, in
much the same sense as the amount of silver in the photographic plates
made by R\"ontgen with x-rays in 1895. In the context of
paleo-detection, this observable has only recently become practically
accessible at the single-event level, through fluorescence imaging of
color centers in dielectric crystals using selective-plane
illumination microscopy
(SPIM)~\cite{Cogswell:2021qlq,Araujo:2025lif,Vladimirov:2024}: in many
materials, vacancies form optically active color centers that
fluoresce. That is, light of a specific wavelength causes each color
center to emit light. Usually the excitation and emission wavelengths
differ at finite temperature due to the Stokes shift. The emitted
fluorescence light can be localized to within about 1/2 of its
wavelength (neglecting details of the microscope optics). We will
refer to a paleo-detector readout based on counting individual color
centers as a ``calorimetric'' readout, in contrast to the
``track-length'' readouts that have dominated the literature so
far. This paper is written in the context of the SPIM technology but
the general findings likely apply to other calorimetric readout
modalities.

The complementarity of tracking and calorimetric readouts has long
been exploited in collider experiments. The general-purpose detectors
at the LHC, CMS and ATLAS~\cite{CMS:2008,ATLAS:2008}, instrument both
a high-resolution tracker that preserves the trajectories of charged
particles and segmented calorimeters that absorb the particles and
measure their total deposited energy. The two readouts give
complementary information --- topology and momentum on the one hand,
total energy on the other --- and combining them is essential for
particle identification. The track-length and vacancy-count readouts
of a paleo-detector are the analogues, at the molecular scale, of the
tracker and the calorimeter at the collider scale, and the same logic
motivates studying their combination here.

In this paper we present the first paleo-detector dark matter
sensitivity analysis based on a calorimetric readout. We perform
full-cascade SRIM simulations~\cite{TRIM,Agarwal:2021} of the dominant
signal and background channels in olivine and extract both track
length and vacancy count event-by-event. We have chosen olivine since
it is considered a good candidate from a geological point of view and
it has been widely used in the paleo-detector literature, allowing for
comparisons with existing results. We propagate the readout-specific
parameters like light yield per vacancy, camera noise, and intrinsic
defect concentration, through to the dark matter sensitivity. We
find that a vacancy-based analysis alone reaches a sensitivity
envelope very similar to state-of-the-art track-based analyses. More
importantly, the combination of the two observables provides an
event-by-event measure of the average $|dE/dx|$ along the track, which
acts as a discriminator of the recoiling nuclear species: since the
neutron-nucleus cross section is approximately constant in nuclear
mass while the dark-matter--nucleus cross section scales as $A^2$,
this discrimination suppresses neutron backgrounds by more than an
order of magnitude at moderate dark matter masses. The resulting
combined-analysis sensitivity for $100\,\mathrm{g}$ of
$1\,\mathrm{Gyr}$ old olivine reaches spin-independent
dark-matter--nucleon cross sections in the $10^{-48}\,\mathrm{cm}^2$
range at WIMP masses of a few tens of GeV, comparable to future liquid
noble gas detectors. We further note that, even at the boundary of the
neutrino fog~\cite{Billard:2013qya,OHare:2021utq}, paleo-detector
experiments would still record $10^3$--$10^5$ dark-matter events ---
the limitation in this regime is systematic, not statistical.

The remainder of this paper is organized as follows. In
section~\ref{sec:tracks} we describe how track length and vacancy
count are extracted from SRIM simulations and discuss the relative
merits of the ``fast'' and ``full cascade'' simulation modes. In
section~\ref{sec:setup} we define the target, exposure, signal model,
and background model used throughout the analysis. We then present
sensitivity results in section~\ref{sec:sensitivity}, comparing
track-only and vacancy-only analyses, examining the impact of
intrinsic defects, demonstrating the species-discrimination power of
the (track length, vacancy count) plane, and presenting the
combined-analysis sensitivity. Section~\ref{sec:fog} discusses the
implications for the neutrino and neutron fog. We comment on practical
large-mass readout in section~\ref{sec:outlook} and conclude in
section~\ref{sec:conclusion}.

\section{Tracks versus vacancies}
\label{sec:tracks}

The premise of paleo-detection is that persistent changes to the
crystal structure encode information about the underlying particle
interaction. A charged particle traversing a solid loses energy
through three distinct channels: ionization, elastic lattice
deformations, and inelastic lattice deformations. Neutral particles
such as neutrinos, neutrons, and dark matter candidates first transfer
their energy to a charged primary knock-on atom (PKA), which then
deposits it through the same channels. For our purposes the relevant
quantity is the inelastic deformation of the lattice, which produces
the persistent vacancy population that constitutes the calorimetric
signal. The other two channels must nevertheless be modeled correctly,
since they govern the path along which vacancies are produced as well as energy loss between collision sites, and
hence the track length.

\begin{figure}
    \centering
    \includegraphics[width=0.5\linewidth]{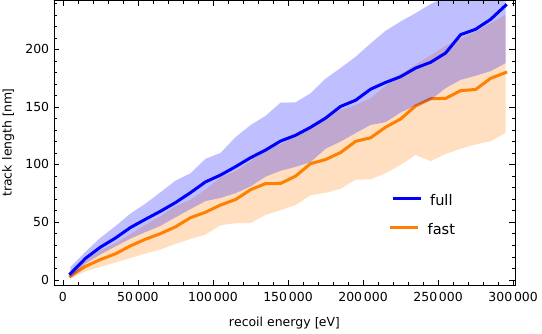}%
    \includegraphics[width=0.5\linewidth]{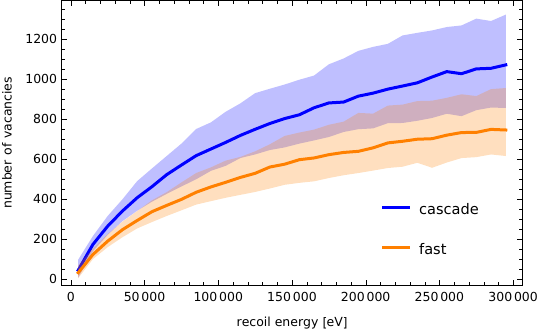}
    \caption{Distribution of track length (left panel) and number of
      vacancies (right panel) as a function of primary recoil energy,
      based on SRIM simulations in full cascade mode (blue) and quick
      mode (orange). The bands indicate the 68\% quantile of the
      distributions. The data shown is for Fe primary knock-on atoms
      in olivine.}
    \label{fig:energy-spectrum}
\end{figure}

The standard tool used in the paleo-detection community for these
calculations is SRIM~\cite{TRIM}. SRIM is a binary-collision Monte
Carlo code that propagates energetic ions through matter, tracking the
electronic and nuclear stopping along each trajectory; for each
primary it returns the full sequence of secondary recoils, the
deposited-energy spectrum, and the resulting populations of vacancies
and interstitials. It has been the de facto standard for
displacement-damage calculations in radiation physics for several
decades. SRIM offers two distinct simulation modes, commonly referred
to as ``quick'' (or ``fast'') and ``full cascade''. In quick mode, defects are generated according to the Kinchin-Pease approximation: the
energy lost in any secondary collision is subtracted from the
primary's energy, and no secondary cascade is launched. In full
cascade mode, by contrast, each secondary recoil generates its own
cascade of subsequent recoils, and so on, until all knock-on atoms
fall below the displacement threshold. In principle, full cascade
should be the more physically faithful of the two, since the actual
displacement physics is hierarchical. In practice, however, the two
modes do not yield consistent results: a long-standing issue,
characterized in detail in Ref.~\cite{Agarwal:2021}, is that full
cascade mode does not correctly account for replacements, in which a
vacancy is filled by a passing interstitial. As a result, full cascade
overpredicts the number of stable vacancies by up to a factor of two
relative to the quick mode, with somewhat smaller differences for the
track length. This effect is illustrated in
Fig.~\ref{fig:energy-spectrum} for both observables, for iron primary
recoils in olivine. We use full cascade simulations as our default
throughout this work, since the resulting event topology is more
realistic; we will return to the impact of the cascade-mode
normalization uncertainty on dark matter sensitivity in
section~\ref{sec:sensitivity}.

The SRIM simulations of olivine were conducted using the surface binding energy, lattice binding energy, and displacement energy from Refs.~\cite{Wang:1999,May:2000} as compiled within Ref.~\cite{Bocchio:2014}. The density of the olivine was assumed to be 3.32\,g\,$\mbox{cm}^{-3}$~\cite{webmineral_olivine}. Primary atoms were simulated with an energy of 1\,MeV---a processing script tracked the energy of this primary as it collided with atoms in the lattice to generate a library of lower-energy primary damage tracks, all rotated to have an initial orientation in the same direction. The primary atom energies of interest for this study are all substantially lower than 1\,MeV, allowing the primary to ``burn in'' and produce damage tracks spanning the 0--300\,keV region of interest.

Near the vacancy production threshold, there is an energy-dependent probability that recoils produce zero vacancies. To compensate for this, vacancy production efficiency curves were generated by running mono-energetic primaries of each species from 0\,eV up until the energy where an efficiency of 99.999\% was achieved, using a 1\,eV step size, typically on the order of a few hundred eV. Separate efficiency curves were generated for the full cascade and fast modes.

There is in general no simple one-to-one correspondence between
vacancy formation and the type of color center created. For instance,
in lithium fluoride, the same anion sublattice supports a family of
related defects, like isolated F centers (anion vacancy $ + e^- $),
di-vacancy $F_2/M$ centers, tri-vacancy $F_3/R$ centers, and their
ionized counterparts $F_2^+ $ and $F_3^+ $, whose absorption and
emission bands span from the UV to the
near-IR~\cite{nahum1967fcenters,baldacchini1996f2f3}. Moreover,
different types of irradiation populate the different types of color
centers at a different rate~\cite{Araujo:2025lif}. Also in diamond we
have many different types of color centers, many stemming from
impurity vacancy combinations, {\it e.g.} the famous $NV$ center, but
also different vacancy centers like the $GR1$ and $ND1$
centers~\cite{zaitsev2001optical}. The sublattice the vacancy sits in
also affects the type of color center formed, for instance in silicon
carbide very different centers result from a missing carbon atom
versus a missing silicon atom~\cite{castelletto2020silicon}. It stands
to reason, that even for other readout modalities the relationship
between vacancy formation and observable signature is far from
trivial, {\it e.g.}, for x-ray-based methods the contrast is a strong
function of the atomic number. This is not only a complication, but
may provide signatures to disentangle the type of radiation that
caused the signature as well as to aid in particle identification. This
topic is rich and deserves further study but goes beyond the scope of
the analysis presented here.  Therefore, for simplicity we assume that
vacancies of all atomic species contribute equally to the calorimetric
signal and each vacancy gives rise to one color center.

\begin{figure}
    \centering
    \includegraphics[width=0.5\linewidth]{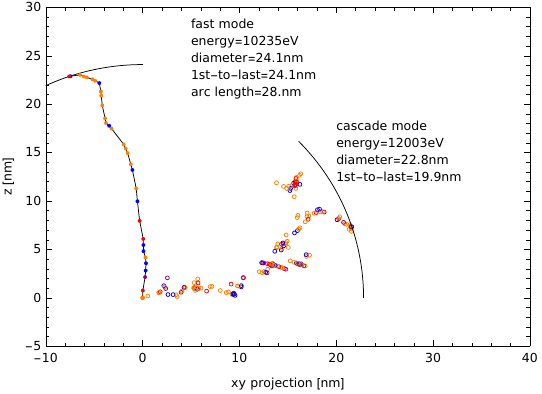}
    \caption{Projection into the $xy$-plane of two events simulated
      with SRIM at approximately $10\,\mathrm{keV}$ primary recoil
      energy. Filled disks show the result in quick mode, open circles
      show the result in full cascade mode. Each colored arc
      corresponds to one of the three track-length definitions
      discussed in the text: Euclidean first-to-last distance, arc
      length along the time-ordered cascade, and diameter
      (largest pairwise vacancy distance). Blue markers denote oxygen vacancies, orange markers denote magnesium vacancies, green markers denote silicon vacancies, and red markers denote iron vacancies.    }
    \label{fig:defs}
\end{figure}

The differences between the two simulation modes of SRIM go
beyond an overall normalization. Figure~\ref{fig:defs} shows two
simulated events of approximately $10\,\mathrm{keV}$ primary recoil
energy: one in quick mode (filled disks) and one in full cascade mode
(open circles). The PKA in both cases is iron, and the colors denote
the atomic number of the recoiling atom along the cascade. The visual
difference is striking: quick mode produces a roughly linear event
with all damage along a single trajectory, whereas full cascade mode
produces a much richer topology with off-axis branches and a more
volume-filling event. This complicates the definition of a single
track-length parameter for either type of simulation. Especially at
lower recoil energies, events tend to become rounder, that is, their
extents along the track direction and transverse to it become more and
more comparable. At the lowest energies we have only a handful of
vacancies that form more of a cloud than a linear track. Moreover,
it is worth pointing out that even the concept of ``track length
resolution'' is not an obvious one: when observing the etch pit of a
fission track of $\sim$20\,$\mu$m length, the resolution required to
see this event as a linear track is not the same as the track
length. The optical microscopes employed in this type of analysis have
an isotropic sub-micron resolution, {\it i.e.} much closer to the
track width than to the length. X-ray techniques such as ptychography or
tomography are more sensitive to the total volume of the disturbance
of the lattice than the linear size of a feature. For very high
resolution techniques like transmission electron microscopy, where
every atom becomes visible, the distinction between track analysis and
vacancy counting vanishes altogether. There is no clear dichotomy
between reading out tracks and vacancy counts, nor is there a general
definition of what type of resolution is actually required to see
tracks of a certain size. These distinctions are conceptually very
useful but hardly reflect experimental realities. With all these
caveats in place, we nonetheless will use these simple categories
going forward since they serve to illustrate the underlying concepts
clearly.

A simple and natural length definition is the Euclidean distance
between the first and last vacancy of the event. This works well in
quick mode, where the cascade is essentially one-dimensional, but it
is less well-defined in full cascade mode, where the transverse
size of the event is not necessarily small compared to its length. A
second option is the arc length obtained by connecting every vacancy
in the time-ordered sequence; again, for full cascade events there is
not one arc connecting all sub-cascades, and a choice would have to be
made as to which sub-cascades to include. A close cousin of the arc length
is the spline-fit technique introduced in Ref.~\cite{Fung:2025cub},
which suffers the same problem for events with sub-cascades. We also note that in real data the distinction of first and last event does not exist. A third
definition that is not predicated on a linear event shape is given by
the diameter of the event; the diameter is defined as the largest pairwise
distance between any two vacancies. The diameter can be computed in
$\mathcal{O}(n\log n)$ time, where $n$ is the number of vacancies, and
is well-defined for any event topology. The three definitions are
illustrated in Fig.~\ref{fig:defs} for both simulation modes. We find
that the spread between definitions is generally smaller than the
difference between full cascade and quick mode, and that the diameter
tracks the other two definitions reasonably well in cases where they
are unambiguous. Since the diameter applies uniformly to both
simulation modes and to events of arbitrary topology, we adopt it as
our default definition of track length throughout this paper. With
these conventions in place, we now turn to the target, exposure, and
background model used in the sensitivity analysis.

\section{Target, exposure, and backgrounds}
\label{sec:setup}

We adopt olivine, $(\mathrm{Mg},\mathrm{Fe})_2\mathrm{SiO}_4$, as the
canonical target throughout this analysis, with the ratio of Mg/Fe being 4.  Olivine is abundant in the
Earth's mantle and in many crustal rocks, has been extensively
characterized in the fission-track-dating
literature~\cite{Wagner:1992,Gleadow:2002}, and contains a mixture of
light and intermediate-mass nuclei, O, Mg, Si, and Fe, which
proves convenient for distinguishing dark matter signals from neutron
backgrounds, as we will discuss in
section~\ref{sec:sensitivity}. Following
Refs.~\cite{Baum:2018tfa,Baum:2023cct,Fung:2025cub} we take a
benchmark exposure of $100\,\mathrm{g}$ of $1\,\mathrm{Gyr}$ old
olivine, corresponding to an integrated
$100\,\mathrm{kt}\cdot\mathrm{yr}$, which is three orders of magnitude
beyond the largest planned liquid noble gas dark matter
experiment~\cite{DARWIN:2016hyl,XLZD:2024}.

For the dark matter signal we assume a spin-independent contact
interaction with nuclei. The WIMP-nucleus cross section is
\begin{equation}
\label{eq:sigma}
\sigma_{\chi N} = \left(\frac{\mu_{\chi N}}{\mu_{\chi n}}\right)^2 A^2\, \sigma_{\chi n}\,,
\end{equation}
where $\sigma_{\chi n}$ is the per-nucleon cross section, $A$ is the
target mass number, and $\mu_{\chi N}$, $\mu_{\chi n}$ are the
WIMP-nucleus and WIMP-nucleon reduced masses. The coherent $A^2$
enhancement is the basis of the species-discrimination argument of
section~\ref{sec:dedx}. The differential recoil rate per unit
target mass is
\begin{equation}
\label{eq:rate}
\frac{dR}{dE_R} = \frac{\rho_\chi\, \sigma_{\chi n}\, A^2}{2\, m_\chi\, \mu_{\chi n}^2}\, F^2(E_R)\, \eta(v_{\rm min})\,,
\end{equation}
where $\rho_\chi$ is the local dark matter density, $m_\chi$ is the
WIMP mass, $F(E_R)$ is the nuclear form factor, and
$\eta(v_{\rm min}) = \int_{v_{\rm min}}^\infty f(v)/v\,dv$ is the mean
inverse speed of the WIMP velocity distribution above the kinematic
threshold $v_{\rm min} = \sqrt{m_N E_R / 2\mu_{\chi N}^2}$. For the
nuclear form factor we use the Helm
parametrization~\cite{Helm:1956,LewinSmith:1996},
\begin{equation}
\label{eq:helm}
F(q) = \frac{3\, j_1(q\, r_n)}{q\, r_n}\, \exp\!\left(-\tfrac{1}{2}(q\, s)^2\right)\,,
\end{equation}
with momentum transfer $q = \sqrt{2\, m_N\, E_R}$, nuclear radius
$r_n^2 = c^2 + \tfrac{7}{3}\pi^2 a^2 - 5 s^2$, and parameters $c =
1.23\, A^{1/3} - 0.60\,\mathrm{fm}$, $a = 0.52\,\mathrm{fm}$, and
surface thickness $s = 0.9\,\mathrm{fm}$ as given in
Ref.~\cite{LewinSmith:1996}. $j_1$ is the order-1 spherical Bessel
function of the first kind. For the WIMP velocity distribution we
adopt the standard truncated Maxwellian with circular speed $v_0 =
220\,\mathrm{km}/\mathrm{s}$, local escape speed $v_{\rm esc} =
544\,\mathrm{km}/\mathrm{s}$~\cite{Smith:2007zzr}, average Earth speed
$v_E = 232\,\mathrm{km}/\mathrm{s}$, and local dark matter density
$\rho_\chi = 0.3\,\mathrm{GeV}/\mathrm{cm}^3$~\cite{Read:2014qva}, in
agreement with Ref.~\cite{Fung:2025cub}. The recoil energy spectrum
from Eq.~\ref{eq:rate} is converted into a distribution in the (track
length, vacancy count) plane through the SRIM simulation pipeline of
section~\ref{sec:tracks}, where each PKA at a given recoil energy is
sampled from the corresponding cascade-by-cascade SRIM output.

Backgrounds fall into two distinct classes: those arising from the
host mineral, and those arising from the imaging readout itself. The
first class is dominated by neutrons produced through spontaneous
fission of $^{238}$U trace impurities and through $(\alpha,n)$
reactions on light nuclei from $\alpha$-emitters in the U and Th decay
chains. We use the {\it paleo-spec}
framework~\cite{Baum:2018tfa,Baum:2021jak} for the neutron recoil
spectrum and propagate it through SRIM in the same way as the
signal. The $^{238}$U concentration is the only normalization
parameter for the neutron background; we adopt $0.1\,\mathrm{ppb}$ as
our canonical value, with a $1\%$ systematic uncertainty consistent
with current rock-chemistry analyses. The $\alpha$-decay of $^{238}$U
into $^{234}$Th produces a recoil nucleus of 72\,keV energy
resulting in a damage track in the $\sim$30\,nm range. We include it
as an additional, well-localized background class. Note, that we
assume that all minerals have been so deep underground, at depths of
$\sim 5,000$\,m or more, that direct muon-induced interactions
can be neglected~\cite{Baum:2018tfa}.

In addition to neutrons, coherent elastic neutrino-nucleus scattering
generates an irreducible background from atmospheric neutrinos and
from solar, diffuse, and galactic supernova neutrinos accumulated over
the exposure period. Solar neutrinos contribute at low recoil energy
through the pp, $^7$Be, pep, $^{13}$N, $^{15}$O, $^{17}$F, $^8$B, and
hep components; the diffuse supernova neutrino background (DSNB) and
the galactic supernova neutrino background (GSNB), the latter
representing $\sim 1\,\mathrm{Gyr}$ of accumulated galactic
core-collapse neutrinos, contribute at intermediate energies;
atmospheric neutrinos contribute at the highest recoil energies, where
they are most relevant for high-mass dark matter searches. We use the
standard fluxes and cross sections as summarized in
Ref.~\cite{Fung:2025cub} and take the numerical values from the {\it
  paleo-spec} framework~\cite{Baum:2018tfa,Baum:2021jak}.

The second background class is specific to the calorimetric readout
and proposed fluorescence microscopy method. In fluorescence
spectroscopy the detectability of a given color center comes down to
the light collection system and the camera noise. Without going into
any technical details we can specify the number of photo-electrons
(p.e.) that can be detected on average per color center, which we vary
between 1 and 10\,p.e. For the camera noise we use the specification
of the Hamamatsu Orca Quest 2.0 camera, that yields 0.3\,p.e. of noise
per pixel.  In addition, naturally occurring lattice defects in olivine
produce a population of optically active color centers that are
indistinguishable from radiation-induced single vacancies. We treat the
intrinsic-defect concentration as a free parameter and study its
impact on sensitivity in section~\ref{sec:sensitivity}; we adopt
$1\,\mathrm{ppb}$ as the canonical value.

\begin{figure}
    \centering
    \includegraphics[width=0.5\linewidth]{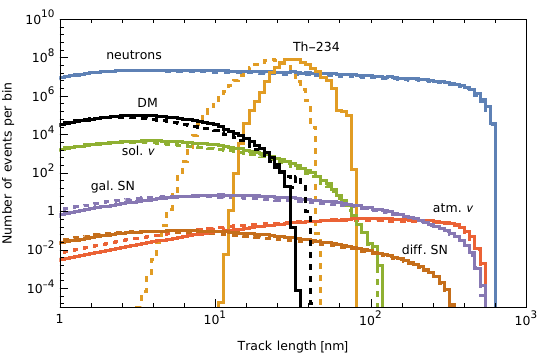}%
    \includegraphics[width=0.5\linewidth]{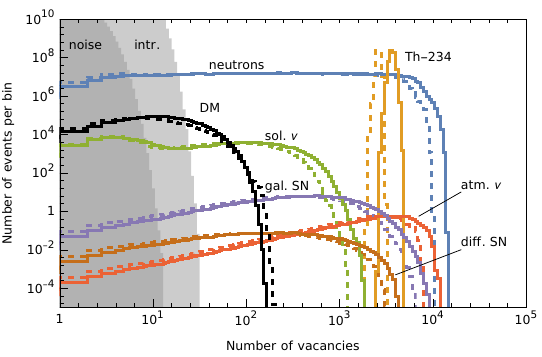}
    \caption{Track length (left panel) and vacancy count (right panel)
      distributions for the dark matter signal at
      $m_\chi=5\,\mathrm{GeV}$ and $\sigma=10^{-45}\,\mathrm{cm}^2$,
      and for each background class, for the $100\,\mathrm{g}\cdot
      1\,\mathrm{Gyr}$ olivine benchmark. Solid curves correspond to
      full cascade simulations, dashed curves to quick-mode
      simulations. The right panel assumes a per-vacancy light yield
      of $10$ photo-electrons. The dark-gray shaded region indicates
      the camera noise floor; the light-gray shaded region indicates
      the contribution from $1\,\mathrm{ppb}$ of intrinsic vacancy
      defects.}
    \label{fig:spec}
\end{figure}

Figure~\ref{fig:spec} shows the resulting per-bin event counts in the
track length (left panel) and vacancy count (right panel) for the dark
matter signal and each background class, computed for the
$100\,\mathrm{g}\cdot 1\,\mathrm{Gyr}$ benchmark exposure. The solid
and dashed curves correspond to full cascade and quick simulation
modes, respectively, illustrating the per-class normalization
differences discussed in section~\ref{sec:tracks}. Several features
are worth noting. Neutrons from $^{238}$U fission dominate the rate at
almost all track lengths and vacancy counts, with the exception of the
lowest-recoil-energy bins where solar neutrinos take over. The daughter of the $^{238}$U $\alpha$-decay, $^{234}$Th produces a sharp, narrow peak in track length around $\sim$30\,nm. The dark matter signal shape overlaps
significantly with all backgrounds in either projection considered
separately: a fact that motivates the combined analysis of
section~\ref{sec:dedx}. The right panel additionally indicates
the calorimetric-readout backgrounds: the camera noise floor at the
lowest photo-electron counts (dark-gray shaded region) and the
contribution from $1\,\mathrm{ppb}$ of intrinsic vacancy defects
(light-gray shaded region).

Table~\ref{tab:syst} summarizes the systematic uncertainties used in
the sensitivity analysis taken from Ref.~\cite{Fung:2025cub}. The
neutrino fluxes are quoted as fractional uncertainties on their
predicted spectra: the solar flux is constrained at the $14\%$ level
by direct measurements, while the DSNB, GSNB, and atmospheric fluxes
are treated as $100\%$ uncertain since they are at most weakly
constrained. The $^{238}$U concentration in any specific olivine
sample can be measured to $1\%$ by destructive chemical analysis, the
mineral age can be determined to $\sim$5\% by independent
geochronology, and the mineral mass is known to better than
$0.01\%$. We note, that in practice these systematic uncertainties
play little role in the analysis since the shapes of the distributions,
in both the track-length and vacancy-count parameters,
distinguish the background from the signal. This, of course, is all
based on simulation, and if in reality the backgrounds are found to be
highly structured this finding will need to be revisited.
\begin{table}[h]
    \centering
    \begin{tabular}{lc}
        \hline
        Parameter $n_i$ & Error $\sigma_i$ \\
        \hline
        $\nu_{\mathrm{solar}}$ flux & $14\%$ \\
        $\nu_{\mathrm{DSNB}}$ flux & $100\%$ \\
        $\nu_{\mathrm{GSNB}}$ flux & $100\%$ \\
        $\nu_{\mathrm{ATM}}$ flux & $100\%$ \\
        $^{238}$U concentration & $1\%$ \\
        Mineral age $t$ & $5\%$ \\
        Mineral mass $m_{\mathrm{olivine}}$ & $0.01\%$ \\
        \hline
    \end{tabular}
    \caption{Systematic uncertainties used in the sensitivity
      analysis, following Ref.~\cite{Fung:2025cub}. Neutrino fluxes
      are quoted as fractional uncertainties on the predicted
      spectra. The mineral parameters are quoted as fractional
      uncertainties on independently-measurable quantities of the host
      olivine sample.}
    \label{tab:syst}
\end{table}

\section{Dark matter sensitivity}
\label{sec:sensitivity}

We compute $90\%$ C.L. exclusion limits on the spin-independent dark
matter--nucleon cross section using a binned Gaussian
likelihood\footnote{A Gaussian likelihood is numerically more
efficient and the event counts even at the limit are in the 1,000s,
hence the underlying Poisson statistics is effectively approximated by
a Gaussian one. We use 1 degree of freedom for determining
the critical value of the test statistic which we assume to follow a
$\chi^2$-distribution.} profiled over the nuisance parameters of
Tab.~\ref{tab:syst}, following the procedure of
Ref.~\cite{Fung:2025cub}. We present the results in four steps that
progressively build up the calorimetric analysis: track-only versus
vacancy-only sensitivity, the impact of intrinsic-defect
contamination, the species-discrimination power of the joint
observable, and the combined-analysis sensitivity. For the track
length we use 100 bins evenly spaced on a logarithmic scale from 1/2
of the track length resolution up to 1000\,nm. We do apply a
resolution function, taken to be a Gaussian smearing in $\log L$ with
width set by the assumed track-length resolution, as described in
Ref.~\cite{Fung:2025cub}. For the
number of vacancies we use 166 bins, where we have bins of width 1
from 1 to 16 and then uniformly on a logarithmic scale up to 500,000
vacancies. We use a Poisson distribution to model the count rate
uncertainty on the photo-electrons. For the combined analysis we use
the same bin sizes but in a two-dimensional grid where events that
fall below the track resolution limit are put into an overflow bin,
{\it i.e.} events in this bin are only binned in vacancy number but
not in track length.

\subsection{Track-only and vacancy-only readout}

\begin{figure}
    \centering
    \includegraphics[width=0.5\linewidth]{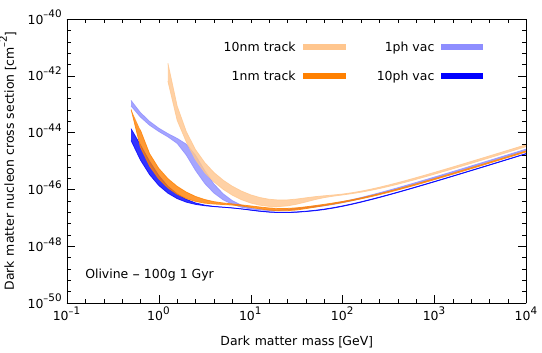}
    \caption{$90\%$ C.L. sensitivity limits on the spin-independent
      dark matter--nucleon cross section as a function of dark matter
      mass, for $100\,\mathrm{g}$ of $1\,\mathrm{Gyr}$ old
      olivine. Track-only analyses are shown for track-length
      resolutions of $1\,\mathrm{nm}$ and $10\,\mathrm{nm}$;
      vacancy-only analyses are shown for per-vacancy light yields of
      $1$ and $10$ photo-electrons. The bands span full cascade
      simulations (lower edge) and quick simulations (upper edge).}
    \label{fig:limits-track-vac}
\end{figure}

Figure~\ref{fig:limits-track-vac} shows the $90\%$ C.L. sensitivity
envelope obtained when only one of the two observables is used. For
the track-only analysis we consider track-length resolutions of
$1\,\mathrm{nm}$ and $10\,\mathrm{nm}$, bracketing what current and
near-future readout techniques are expected to
achieve~\cite{Baum:2023cct}. For the vacancy-only analysis we vary the
per-vacancy light yield between $1$ and $10$ photo-electrons,
bracketing the range expected for existing and improved imaging
readouts~\cite{Cogswell:2021qlq,Araujo:2025lif}. The bands on each
curve indicate the spread between full cascade and quick simulation
modes, which is largest at low dark matter mass and amounts to at most
a factor of a few in the cross section.

The principal observation from Fig.~\ref{fig:limits-track-vac} is that
the vacancy-only and track-only analyses reach very similar
sensitivities across the entire mass range. At dark matter masses
above $\sim$10\,GeV both readouts approach the
$10^{-47}\,\mathrm{cm}^2$ scale. The two
readouts differ only in the low-mass tail, where the small number of
vacancies per signal event makes the calorimetric channel slightly
less sensitive than a high-resolution track-length readout. This
near-equivalence is an interesting and important finding; it also
highlights that the real performance of the readout technology will be
the decisive factor for low-mass performance.

\subsection{Intrinsic-defect contamination}

\begin{figure}
    \centering
    \includegraphics[width=0.5\linewidth]{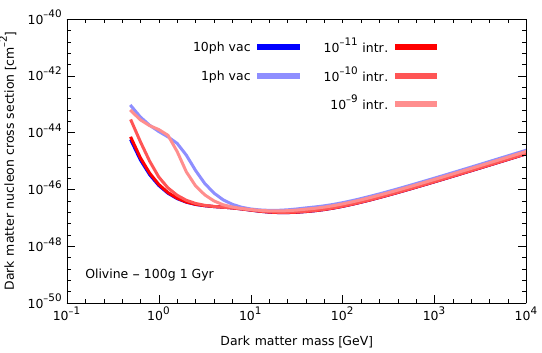}
    \caption{$90\%$ C.L. sensitivity limits on the spin-independent
      dark matter--nucleon cross section as a function of dark matter
      mass, for $100\,\mathrm{g}$ of $1\,\mathrm{Gyr}$ old
      olivine. The vacancy-only sensitivity is shown for an assumed
      light yield of $10$ photo-electrons per vacancy and
      intrinsic-defect concentrations of $0.01$, $0.1$, and
      $1\,\mathrm{ppb}$ (red, from darkest to lightest). For
      comparison, the defect-free results for $1$ and $10$
      photo-electrons per vacancy are also shown (blue, light and
      dark).}
    \label{fig:limits-intr}
\end{figure}

Naturally occurring lattice defects in olivine are an irreducible
background specific to the calorimetric readout. To quantify their
impact we vary the intrinsic-defect concentration over the range
$0.01$ to $1\,\mathrm{ppb}$ while holding all other
parameters at their canonical values. The result is shown in
Fig.~\ref{fig:limits-intr}.

We find that the high-mass sensitivity is essentially unaffected by
intrinsic defects: signal events at high recoil energy produce
hundreds of vacancies, far above the few-vacancy contamination floor
expected even at the highest defect concentrations considered
here. The low-mass tail, however, is sensitive to the defect
concentration. Above $\sim 1\,\mathrm{ppb}$ the calorimetric sensitivity to dark
matter masses below $\sim$5\,GeV degrades by an order of magnitude or
more; below $\sim 0.1\,\mathrm{ppb}$ the impact is negligible. We therefore
identify $\sim 1\,\mathrm{ppb}$ as the optical active defect concentration above which
the low-mass calorimetric channel is significantly compromised, and as
a benchmark for the required olivine purity in any practical
implementation.

\subsection{Species discrimination in the (track length, vacancy count) plane}
\label{sec:dedx}

The marginal distributions of Fig.~\ref{fig:spec} obscure information
that is preserved in the joint distribution. The neutron-nucleus cross
section is approximately flat in nuclear mass, and so neutron-induced
recoils populate all species in olivine roughly in proportion to their
abundance. The spin-independent dark matter--nucleus cross section, by
contrast, scales as $A^2$ in the Helm form factor regime, and so dark
matter recoils concentrate strongly on iron, the heaviest abundant
nucleus in olivine. At a given recoil energy, heavier nuclei have lower velocity, so a
larger fraction of their energy goes into nuclear rather than
electronic stopping; their nuclear stopping power per unit length is
also larger. The combined effect is that heavier PKAs produce shorter
tracks and more vacancies. The pair (track length, vacancy count) thus
acts as a per-event proxy for the specific energy loss $|dE/dx|$ and
consequently for the recoiling species.

\begin{figure}
    \centering
    \includegraphics[width=0.5\linewidth]{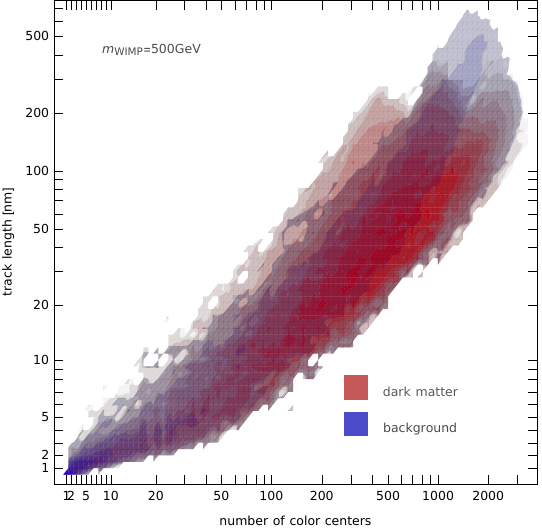}
    \caption{Joint two-dimensional histograms in the (vacancy count, track length) plane for the neutron background (blue) and a dark matter signal at $m_\mathrm{WIMP}=500\,\mathrm{GeV}$ (red), for the $100\,\mathrm{g}\cdot 1\,\mathrm{Gyr}$ olivine benchmark. Color intensity represents the per-bin event density of each distribution on a logarithmic scale, with darker shading indicating higher density.}
    \label{fig:dedx}
\end{figure}

This expectation is borne out in Fig.~\ref{fig:dedx}, which shows the
joint two-dimensional distribution of track length and vacancy count
for the neutron background and for a $500\,\mathrm{GeV}$ mass dark
matter signal. The two populations overlap in either projection but
separate clearly in the joint plane: the dark matter signal occupies
the region of higher vacancy count and shorter tracks, corresponding
to recoils on iron, while the neutron background extends to the
lower-vacancy-count, longer-track region populated by recoils on
oxygen, magnesium, and silicon. The same separation persists at other
masses but practically becomes less relevant for low recoil energies
since here the signal in any case is just a handful of vacancies and
a track length can hardly be defined. This
species-discrimination power is the conceptual basis for the
calorimetric advantage: it suppresses the dominant neutron background
at the per-event level, without any reliance on its absolute
normalization. Whether the same plane can be used to discriminate
between different dark matter--nucleon couplings, beyond the
spin-independent contact interaction considered here, is an
interesting question that we leave for future work. It also is worth
noting that minerals that are composed of elements covering a large
range in atomic masses will exhibit this effect to a larger degree
than minerals that are chemically pure, {\it e.g.} diamond. Olivine
spans roughly a range of 3.5 in atomic masses, but for instance zircon
spans a range of 5.7 and barite a range of 8.5; this adds a novel
selection criterion for target minerals for paleo-detection. As
mentioned in section~\ref{sec:tracks}, in reality several species of color
centers will be produced and likely can be distinguished
spectroscopically, enhancing the particle identification.  Also using track length as a single descriptor of the spatial morphology of the event is probably not the best way to characterize the information contained in the event and other more complex measures of morphology could be employed. 

\subsection{Combined analysis}

\begin{figure}
    \centering
    \includegraphics[width=0.5\linewidth]{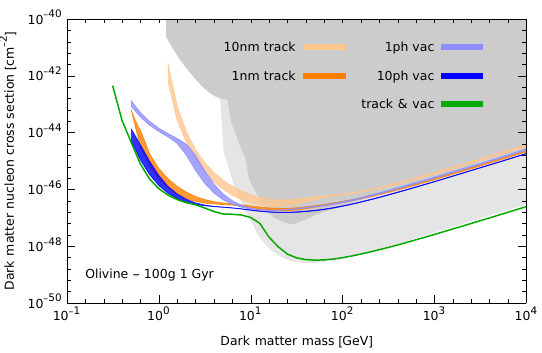}
    \caption{$90\%$ C.L. sensitivity limits on the spin-independent
      dark matter--nucleon cross section as a function of dark matter
      mass, for $100\,\mathrm{g}$ of $1\,\mathrm{Gyr}$ old
      olivine. The green curve shows the combined analysis using the
      joint (track length, vacancy count) likelihood; the orange and
      blue curves show the track-only and vacancy-only envelopes from
      Fig.~\ref{fig:limits-track-vac}. The dark-gray and light-gray
      shaded regions indicate the current and projected reach of
      liquid noble gas direct-detection experiments. All results use
      full cascade simulations.}
    \label{fig:limits-all}
\end{figure}

Figure~\ref{fig:limits-all} shows the combined-analysis sensitivity
(green curve), computed using the joint likelihood in the (track
length, vacancy count) plane, alongside the track-only and
vacancy-only envelopes from Fig.~\ref{fig:limits-track-vac}. Since the analysis is based on the fill two-dimensional likelihood not cuts are used to reject neutron backgrounds. The
combined analysis reaches $\sigma \sim 10^{-48}\,\mathrm{cm}^2$ at
$m_\chi$ of a few tens of GeV, roughly an order of magnitude below
either single-observable analysis at the optimum mass. The improvement
comes from the species-discrimination argument of
section~\ref{sec:dedx}: the joint fit suppresses neutron contamination
at the per-event level rather than relying solely on its absolute
normalization.

The dark-gray and light-gray shaded regions in
Fig.~\ref{fig:limits-all} indicate the current and projected reach of
liquid noble gas dark matter
experiments~\cite{LZ:2024zvo,XLZD:2024,DARWIN:2016hyl}. The
combined-analysis envelope reaches sensitivity comparable to the
projected reach of XLZD-class detectors at moderate dark matter
masses, with somewhat better low-mass reach where the lower energy
threshold of crystal targets and the long exposure period of
paleo-detectors are both decisive advantages. At high mass, $m_\chi\gtrsim 1\,\mathrm{TeV}$, the
combined analysis remains comparable to the projected liquid noble gas
reach. The achievable sensitivity is ultimately set by the irreducible
neutrino background, which we discuss in the next section.

\section{The neutrino and neutron fog}
\label{sec:fog}

The achievable sensitivity of any rare-event search is ultimately
limited by irreducible backgrounds, which for direct dark matter
detection usually means coherent elastic neutrino-nucleus scattering. The
corresponding sensitivity floor is conventionally referred to as the
{\it neutrino fog}~\cite{Billard:2013qya,OHare:2021utq}; for
paleo-detectors the long exposure period also makes neutron
backgrounds irreducible, and we will refer to the joint structure as
the {\it neutrino and neutron fog}. Following
Ref.~\cite{OHare:2021utq} we quantify the depth of the fog by the
slope of the discovery cross section with exposure,
\begin{equation}
\label{eq:n}
n = \left(\frac{\mathrm{d}\ln\sigma}{\mathrm{d}\ln N}\right)^{-1}\,.
\end{equation}
For a background-free experiment $n=1$, corresponding to $\sigma_{\rm
  disc}\propto 1/N$. For an experiment dominated by an irreducible
background of perfectly known normalization, Asimov statistics give
$n=2$. In the limit where systematic uncertainties on the background
dominate over statistical fluctuations, $n\rightarrow\infty$ and
additional exposure brings no further gain. The conventional
definition of the fog boundary is the contour $n=2$.

\begin{figure}
    \centering
    \includegraphics[width=1.0\linewidth]{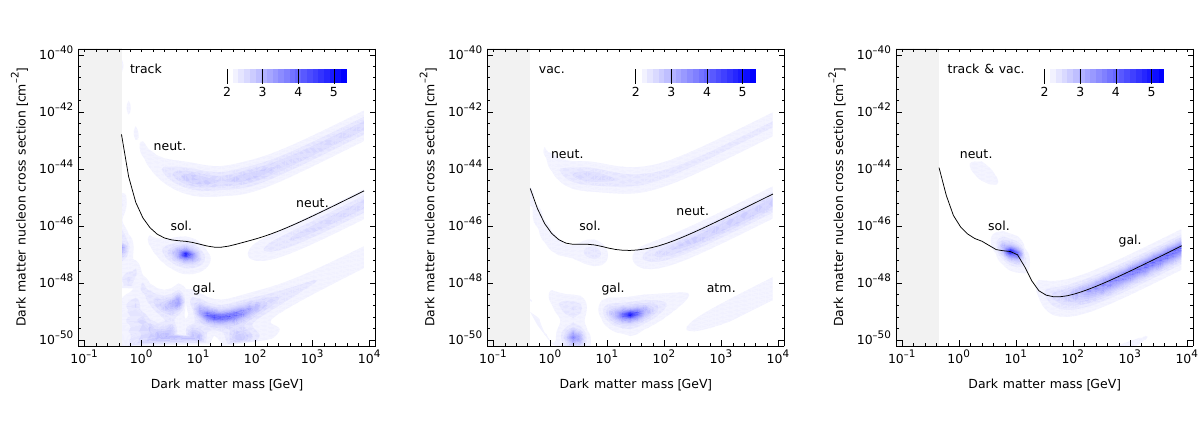}
    \caption{Neutrino and neutron fog for the $100\,\mathrm{g}\cdot
      1\,\mathrm{Gyr}$ olivine benchmark, for the track-only (left
      panel), vacancy-only (middle panel), and combined (right panel)
      analyses. Colored regions show contours of the fog parameter
      $n=(\mathrm{d}\ln\sigma/\mathrm{d}\ln N)^{-1}$ (Eq.~\ref{eq:n}),
      and the black lines mark the corresponding $90\%$
      C.L. sensitivities. The labels identify the dominant background
      behind each island: neutrons (``neut.''), solar neutrinos
      (``sol.''), galactic supernova neutrinos (``gal.''), atmospheric
      neutrinos (``atm.''), and the diffuse supernova neutrino
      background (``diff.''). The transition to $n=1$ for cross
      sections in the $10^{-42}$--$10^{-40}\,\mathrm{cm}^2$ range
      corresponds to the background-free regime where statistical
      scaling becomes linear in exposure.}
    \label{fig:fog}
\end{figure}

Figure~\ref{fig:fog} shows the fog structure for our
$100\,\mathrm{g}\cdot 1\,\mathrm{Gyr}$ olivine benchmark, in three
configurations: a track-only analysis (left panel), a vacancy-only
analysis (middle panel), and the combined analysis of
section~\ref{sec:sensitivity} (right panel). In each panel the colored
regions are contours of $n$ in the (dark matter mass, cross section)
plane, and the black line is the corresponding $90\%$
C.L. sensitivity. Several distinct islands are visible, each driven by
a specific background: neutrons (``neut.''), solar neutrinos
(``sol.''), galactic supernova neutrinos (``gal.''), atmospheric
neutrinos (``atm.''), and the diffuse supernova neutrino background
(``diff.''). The transition to $n=1$, where the experiment becomes effectively
background-free and the discovery cross section scales linearly with
exposure, occurs at cross sections in the $10^{-42}$ to
$10^{-40}\,\mathrm{cm}^2$ range and is not shown. Realistic
paleo-detector targets lie well below this transition, so any
practical search operates inside the fog, be it neutrons or
neutrinos.

Comparing the three panels, the most striking feature is the recession
of the fog when going from single-observable to combined analysis. The
neutron-fog island, dominant in both single-observable panels, is
largely suppressed in the combined panel, reflecting the
species-discrimination power discussed in section~\ref{sec:dedx}. The
combined analysis pierces the floor at moderate dark matter masses,
where the dominant background after combination is a much smaller
neutrino contribution. At low and high masses the floor is instead set
by neutrino backgrounds, from solar and atmospheric neutrinos
respectively, and is essentially unchanged from the single-observable
case --- for the specific set of elements, atomic masses and neutron numbers the specific energy loss for neutrino-induced and dark matter-induced recoils is too similar to be resolved even with the combination of track length and vacancy count. 

A further noteworthy feature is the absolute event count at the
fog-implied limit. Where conventional direct detection experiments
encounter the neutrino floor at expected signal counts of order tens
of events, our $100\,\mathrm{g}\cdot 1\,\mathrm{Gyr}$ olivine
benchmark would record $10^3$--$10^5$ dark matter events at
the same boundary, depending on mass. The limitation in this regime is
therefore systematic rather than statistical. Many of the
systematic uncertainties will come from the geological history of the sample.

\section{Practical readout considerations}
\label{sec:outlook}

The sensitivity benchmark assumed throughout this paper is
$100\,\mathrm{g}$ of olivine, as a sample of that size contains about
$10^9$ events from neutron and neutrino interactions, the dominant
contribution coming from neutron-induced recoils accumulated over the
$1\,\mathrm{Gyr}$ exposure. Reading out this volume at the 1\,nm
resolution required for individual event reconstruction using the track
length as a signature would generate a data volume of about
100 zettabytes ($10^{23}$ Bytes), comparable to or exceeding that of
the high-luminosity LHC at the detector
level~\cite{cms2018phase2daq}. This indicates that a single-stage
readout at nanometer resolution across the full $100\,\mathrm{g}$
sample would currently be very difficult, if not impossible, simply
based on data volumes alone.

\begin{figure}
  
    \includegraphics[width=0.4\linewidth]{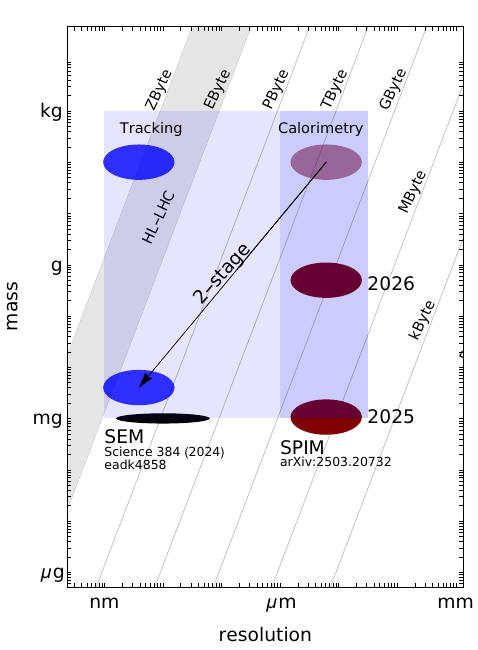}
    \caption{Mass-versus-resolution landscape for paleo-detector
      readout. SPIM (selective-plane illumination microscopy) reaches
      gram-scale samples at micron resolution; SEM (scanning electron
      microscopy) reaches milligram-scale samples at nanometer
      resolution. The two-stage approach (arrow) composes the two to
      reach the $100\,\mathrm{g}$ benchmark of this paper at nanometer
      resolution. Reference markers indicate current capabilities of
      SPIM~\cite{Araujo:2025lif} and SEM~\cite{ShapsonCoe:2024}; the
      2026 SPIM marker corresponds to a recent $0.5\,\mathrm{cm}^3$
      LiF measurement in the Huber lab~\cite{hedges_talk_2026}.
      Diagonal lines indicate constant data volumes, with the
      high-luminosity LHC band shown for
      comparison~\cite{cms2018phase2daq}.}
    \label{fig:stages}
\end{figure}

The key observation is that the high-resolution readout does not need
to be applied to the entire sample volume; only to the small fraction
occupied by candidate events. This motivates a two-stage approach. In
the first stage, the entire sample is scanned at coarse, micron-scale
resolution using selective-plane illumination
microscopy~\cite{Vladimirov:2024,Araujo:2025lif}. Recent measurements
in the Huber lab have demonstrated SPIM imaging of LiF samples up to
$0.5\,\mathrm{cm}^3$ in volume~\cite{hedges_talk_2026}, indicated by the
2026 SPIM marker in Fig.~\ref{fig:stages}. Regions in which the SPIM scan identifies one or more color
centers are flagged as candidate events, and from Fig.~\ref{fig:spec}
together with the optical resolution of the type of SPIM used for
paleo-detection~\cite{Araujo:2025lif} we can infer that each event
would be contained in a $1\,\mu\mathrm{m}^3$ volume. Thus the total
volume for the high-resolution scan is about
$10^9\,\mu\mathrm{m}^3=0.001\,\mathrm{cm}^3$. Given the density of
olivine, the full volume is about $30\,\mathrm{cm}^3$, thus only 1 part
in 30,000 of the volume requires a high-resolution readout.
In the second stage, only those flagged
regions of interest are imaged at the nm-scale using scanning electron
microscopy (SEM), which resolves individual color centers but at a
much smaller per-volume rate. The first stage prefilters the volume;
the second stage delivers the resolution.

Recent results in connectomics show that a petascale SEM analysis of
$\sim 1\,\mathrm{mm}^3$ of human cortex is
possible~\cite{ShapsonCoe:2024}, generating $\sim 1.4$\,petabytes of
imaging data and reconstructing $\sim 150$\,million synapses across
$57{,}000$ cells in a fully-automated pipeline. Modern multi-beam SEMs
can image $\mathrm{mm}^3$-sized regions at nanometer
resolution. The time required is still significant; in this example
326 days of scanning were required~\cite{ShapsonCoe:2024}, and the
corresponding data rates have been demonstrated to be tractable. The
high-resolution scan volume required for a $100\,\mathrm{g}$
paleo-detector readout is comparable to this, provided that
the SPIM stage prefilters the sample efficiently. This simple argument
of course does not address how to isolate the $10^9$ ROIs within
the 100\,g sample and how to make them accessible for SEM. In the
connectomics example the tissue was physically sliced into 30\,nm-thick
sections and each section was then scanned in full under the SEM,
taking full advantage of modern multi-beam SEM technology. It does,
however, show that raw scanning and data-handling capacity exists that
is close to what is required for a full two-stage dual-readout of a
100\,g sample.

Figure~\ref{fig:stages} sketches the relevant mass-versus-resolution
landscape. The single-stage SPIM and SEM technologies span
complementary regions of the (mass, resolution) plane: SPIM reaches
gram-scale samples at micron resolution, SEM reaches milligram-scale
samples at nanometer resolution. Diagonal lines in the figure indicate
contours of constant data volume, with the band corresponding to
high-luminosity LHC data rates~\cite{cms2018phase2daq} shown for
reference. The two-stage approach (arrow) bridges the gap, reaching
the $100\,\mathrm{g}$ benchmark of this paper at the resolution
required for individual color-center localization. The data volumes
involved are formidable but not unprecedented; the analysis pattern,
a small number of high-resolution windows distributed across a much
larger coarse-scan volume, is theoretically well-matched to automated
pipelines. An actual working implementation for paleo-detection, however,
remains to be developed.

\section{Conclusion}
\label{sec:conclusion}

We have presented the first paleo-detector dark matter sensitivity
analysis based on a calorimetric readout, in which the number of
stable lattice vacancies produced by each nuclear recoil is used as a
per-event observable complementary to the track length. Using
full-cascade SRIM simulations of the dominant signal and background
channels in olivine, neutrons from $^{238}$U fission, $^{234}$Th
$\alpha$-decay recoil, atmospheric and solar neutrinos, and diffuse
and galactic supernova neutrinos, we have computed the expected
sensitivity for a $100\,\mathrm{g}\cdot 1\,\mathrm{Gyr}$ exposure. New
parameters introduced by the calorimetric readout like per-vacancy
light yield, camera noise, and intrinsic-defect concentration have
been studied. We find that the differences between full cascade and
quick SRIM modes lead to at most a factor-of-a-few uncertainty in the
cross-section limits and do not qualitatively change any of our
conclusions. For a real experiment these discrepancies will need to be
resolved, of course, and any simulation will need to be calibrated
against real data.

The principal finding is that, for the benchmark exposure considered
here, a vacancy-only readout reaches a sensitivity envelope very
similar to that of state-of-the-art track-only analyses. More
importantly, the combination of the two observables provides an
event-by-event proxy for $|dE/dx|$ and hence for the recoiling nuclear
species. Because the neutron-nucleus cross section is approximately
flat in $A$ while the dark matter--nucleus cross section scales as
$A^2$, this discrimination suppresses the dominant neutron background
by more than an order of magnitude at moderate dark matter masses. The
combined-analysis sensitivity is comparable to the projected reach of
XLZD-class liquid noble gas detectors, with somewhat better low-mass
reach. Even at the
boundary of the neutrino and neutron fog, paleo-detector experiments
would record $10^3$--$10^5$ dark matter events: the limitation in
this regime will be systematics, mostly stemming from the geological
history of the sample.

Reading out a $100\,\mathrm{g}$ olivine sample at the required
resolution is a formidable task. A two-stage approach combining
selective-plane illumination microscopy with scanning electron
microscopy on flagged regions of interest brings the data volumes to a
level comparable with petascale connectomics workflows. Recent
progress in connectomics gives rise to the hope that a two-stage
readout scheme may be approaching feasibility. Importantly, readout of
100\,g of material is currently feasible using calorimetry alone; the
resulting sensitivity to dark matter of a purely calorimetric readout
is comparable to that of a track-based paleo-detector. A dual
calorimetric/track paleo-detector offers a complementary path to the
conventional liquid noble gas approach, achieving competitive dark
matter sensitivity through long exposure rather than large
instantaneous mass.

\section*{Acknowledgments}

This work was supported by the U.S. Department of Energy National
Nuclear Security Administration's Enabling Capabilities in Technology
(TecH) Consortium under award number DE-NA0004197.  It was also
supported by the U.S. Department of Energy Office of Science under
award number DE-SC0020262. It was also supported by a National Science
Foundation Growing Convergence Research award number 2428507.

\bibliography{biblio}

\begin{thebibliography}{43}%
\makeatletter
\providecommand \@ifxundefined [1]{%
 \@ifx{#1\undefined}
}%
\providecommand \@ifnum [1]{%
 \ifnum #1\expandafter \@firstoftwo
 \else \expandafter \@secondoftwo
 \fi
}%
\providecommand \@ifx [1]{%
 \ifx #1\expandafter \@firstoftwo
 \else \expandafter \@secondoftwo
 \fi
}%
\providecommand \natexlab [1]{#1}%
\providecommand \enquote  [1]{``#1''}%
\providecommand \bibnamefont  [1]{#1}%
\providecommand \bibfnamefont [1]{#1}%
\providecommand \citenamefont [1]{#1}%
\providecommand \href@noop [0]{\@secondoftwo}%
\providecommand \href [0]{\begingroup \@sanitize@url \@href}%
\providecommand \@href[1]{\@@startlink{#1}\@@href}%
\providecommand \@@href[1]{\endgroup#1\@@endlink}%
\providecommand \@sanitize@url [0]{\catcode `\\12\catcode `\$12\catcode
  `\&12\catcode `\#12\catcode `\^12\catcode `\_12\catcode `\%12\relax}%
\providecommand \@@startlink[1]{}%
\providecommand \@@endlink[0]{}%
\providecommand \url  [0]{\begingroup\@sanitize@url \@url }%
\providecommand \@url [1]{\endgroup\@href {#1}{\urlprefix }}%
\providecommand \urlprefix  [0]{URL }%
\providecommand \Eprint [0]{\href }%
\providecommand \doibase [0]{http://dx.doi.org/}%
\providecommand \selectlanguage [0]{\@gobble}%
\providecommand \bibinfo  [0]{\@secondoftwo}%
\providecommand \bibfield  [0]{\@secondoftwo}%
\providecommand \translation [1]{[#1]}%
\providecommand \BibitemOpen [0]{}%
\providecommand \bibitemStop [0]{}%
\providecommand \bibitemNoStop [0]{.\EOS\space}%
\providecommand \EOS [0]{\spacefactor3000\relax}%
\providecommand \BibitemShut  [1]{\csname bibitem#1\endcsname}%
\let\auto@bib@innerbib\@empty
%</preamble>
\bibitem [{\citenamefont {Price}\ and\ \citenamefont
  {Walker}(1963)}]{price1963fossil}%
  \BibitemOpen
  \bibfield  {author} {\bibinfo {author} {\bibfnamefont {P.~B.}\ \bibnamefont
  {Price}}\ and\ \bibinfo {author} {\bibfnamefont {R.~M.}\ \bibnamefont
  {Walker}},\ }\bibfield  {title} {\enquote {\bibinfo {title} {Fossil tracks of
  charged particles in mica and the age of minerals},}\ }\href {\doibase
  10.1029/JZ068i016p04847} {\bibfield  {journal} {\bibinfo  {journal} {Journal
  of Geophysical Research}\ }\textbf {\bibinfo {volume} {68}},\ \bibinfo
  {pages} {4847--4862} (\bibinfo {year} {1963})}\BibitemShut {NoStop}%
\bibitem [{\citenamefont {Wagner}\ and\ \citenamefont {{Van den
  haute}}(1992)}]{Wagner:1992}%
  \BibitemOpen
  \bibfield  {author} {\bibinfo {author} {\bibfnamefont {G.~A.}\ \bibnamefont
  {Wagner}}\ and\ \bibinfo {author} {\bibfnamefont {P.}~\bibnamefont {{Van den
  haute}}},\ }\href@noop {} {\emph {\bibinfo {title} {{Fission Track
  Dating}}}}\ (\bibinfo  {publisher} {Ferdinand Enke Verlag},\ \bibinfo
  {address} {Stuttgart},\ \bibinfo {year} {1992})\BibitemShut {NoStop}%
\bibitem [{\citenamefont {Gleadow}\ \emph {et~al.}(2002)\citenamefont
  {Gleadow}, \citenamefont {Belton}, \citenamefont {Kohn},\ and\ \citenamefont
  {Brown}}]{Gleadow:2002}%
  \BibitemOpen
  \bibfield  {author} {\bibinfo {author} {\bibfnamefont {A.~J.~W.}\
  \bibnamefont {Gleadow}}, \bibinfo {author} {\bibfnamefont {D.~X.}\
  \bibnamefont {Belton}}, \bibinfo {author} {\bibfnamefont {B.~P.}\
  \bibnamefont {Kohn}}, \ and\ \bibinfo {author} {\bibfnamefont {R.~W.}\
  \bibnamefont {Brown}},\ }\bibfield  {title} {\enquote {\bibinfo {title}
  {{Fission track dating of phosphate minerals and the thermochronology of
  apatite}},}\ }\href@noop {} {\bibfield  {journal} {\bibinfo  {journal}
  {Reviews in Mineralogy and Geochemistry}\ }\textbf {\bibinfo {volume} {48}},\
  \bibinfo {pages} {579--630} (\bibinfo {year} {2002})}\BibitemShut {NoStop}%
\bibitem [{\citenamefont {Fleischer}\ \emph {et~al.}(1969)\citenamefont
  {Fleischer}, \citenamefont {Hart}, \citenamefont {Jacobs}, \citenamefont
  {Price}, \citenamefont {Schwarz},\ and\ \citenamefont
  {Aumento}}]{fleischer1969monopoles}%
  \BibitemOpen
  \bibfield  {author} {\bibinfo {author} {\bibfnamefont {R.~L.}\ \bibnamefont
  {Fleischer}}, \bibinfo {author} {\bibfnamefont {Jr.}\ \bibnamefont {Hart},
  \bibfnamefont {H.~R.}}, \bibinfo {author} {\bibfnamefont {I.~S.}\
  \bibnamefont {Jacobs}}, \bibinfo {author} {\bibfnamefont {P.~B.}\
  \bibnamefont {Price}}, \bibinfo {author} {\bibfnamefont {W.~M.}\ \bibnamefont
  {Schwarz}}, \ and\ \bibinfo {author} {\bibfnamefont {F.}~\bibnamefont
  {Aumento}},\ }\bibfield  {title} {\enquote {\bibinfo {title} {Search for
  magnetic monopoles in deep ocean deposits},}\ }\href {\doibase
  10.1103/PhysRev.184.1393} {\bibfield  {journal} {\bibinfo  {journal}
  {Physical Review}\ }\textbf {\bibinfo {volume} {184}},\ \bibinfo {pages}
  {1393--1397} (\bibinfo {year} {1969})}\BibitemShut {NoStop}%
\bibitem [{\citenamefont {Price}\ and\ \citenamefont
  {Salamon}(1986)}]{price1986supermassive}%
  \BibitemOpen
  \bibfield  {author} {\bibinfo {author} {\bibfnamefont {P.~B.}\ \bibnamefont
  {Price}}\ and\ \bibinfo {author} {\bibfnamefont {M.~H.}\ \bibnamefont
  {Salamon}},\ }\bibfield  {title} {\enquote {\bibinfo {title} {Search for
  supermassive magnetic monopoles using mica crystals},}\ }\href {\doibase
  10.1103/PhysRevLett.56.1226} {\bibfield  {journal} {\bibinfo  {journal}
  {Physical Review Letters}\ }\textbf {\bibinfo {volume} {56}},\ \bibinfo
  {pages} {1226--1229} (\bibinfo {year} {1986})}\BibitemShut {NoStop}%
\bibitem [{\citenamefont {Snowden-Ifft}\ \emph {et~al.}(1995)\citenamefont
  {Snowden-Ifft}, \citenamefont {Chan},\ and\ \citenamefont
  {Frenkel}}]{SnowdenIfft:1995}%
  \BibitemOpen
  \bibfield  {author} {\bibinfo {author} {\bibfnamefont {D.~P.}\ \bibnamefont
  {Snowden-Ifft}}, \bibinfo {author} {\bibfnamefont {M.~K.~Y.}\ \bibnamefont
  {Chan}}, \ and\ \bibinfo {author} {\bibfnamefont {R.}~\bibnamefont
  {Frenkel}},\ }\bibfield  {title} {\enquote {\bibinfo {title} {{Compositional
  dependence of the formation of nuclear tracks in muscovite mica}},}\
  }\href@noop {} {\bibfield  {journal} {\bibinfo  {journal} {Phys. Rev. Lett.}\
  }\textbf {\bibinfo {volume} {74}},\ \bibinfo {pages} {4133--4136} (\bibinfo
  {year} {1995})}\BibitemShut {NoStop}%
\bibitem [{\citenamefont {Drukier}\ \emph {et~al.}(2019)\citenamefont
  {Drukier}, \citenamefont {Baum}, \citenamefont {Freese}, \citenamefont
  {G{\'o}rski},\ and\ \citenamefont {Stengel}}]{Drukier:2018pdt}%
  \BibitemOpen
  \bibfield  {author} {\bibinfo {author} {\bibfnamefont {Andrzej~K.}\
  \bibnamefont {Drukier}}, \bibinfo {author} {\bibfnamefont {Sebastian}\
  \bibnamefont {Baum}}, \bibinfo {author} {\bibfnamefont {Katherine}\
  \bibnamefont {Freese}}, \bibinfo {author} {\bibfnamefont {Maciej}\
  \bibnamefont {G{\'o}rski}}, \ and\ \bibinfo {author} {\bibfnamefont
  {Patrick}\ \bibnamefont {Stengel}},\ }\bibfield  {title} {\enquote {\bibinfo
  {title} {{Paleo-detectors: Searching for Dark Matter with Ancient
  Minerals}},}\ }\href {\doibase 10.1103/PhysRevD.99.043014} {\bibfield
  {journal} {\bibinfo  {journal} {Phys. Rev. D}\ }\textbf {\bibinfo {volume}
  {99}},\ \bibinfo {pages} {043014} (\bibinfo {year} {2019})},\ \Eprint
  {http://arxiv.org/abs/1811.06844} {arXiv:1811.06844 [astro-ph.CO]}
  \BibitemShut {NoStop}%
\bibitem [{\citenamefont {Baum}\ \emph {et~al.}(2020)\citenamefont {Baum},
  \citenamefont {Drukier}, \citenamefont {Freese}, \citenamefont {G{\'o}rski},\
  and\ \citenamefont {Stengel}}]{Baum:2018tfa}%
  \BibitemOpen
  \bibfield  {author} {\bibinfo {author} {\bibfnamefont {Sebastian}\
  \bibnamefont {Baum}}, \bibinfo {author} {\bibfnamefont {Andrzej~K.}\
  \bibnamefont {Drukier}}, \bibinfo {author} {\bibfnamefont {Katherine}\
  \bibnamefont {Freese}}, \bibinfo {author} {\bibfnamefont {Maciej}\
  \bibnamefont {G{\'o}rski}}, \ and\ \bibinfo {author} {\bibfnamefont
  {Patrick}\ \bibnamefont {Stengel}},\ }\bibfield  {title} {\enquote {\bibinfo
  {title} {{Searching for Dark Matter with Paleo-Detectors}},}\ }\href
  {\doibase 10.1016/j.physletb.2020.135325} {\bibfield  {journal} {\bibinfo
  {journal} {Phys. Lett. B}\ }\textbf {\bibinfo {volume} {803}},\ \bibinfo
  {pages} {135325} (\bibinfo {year} {2020})},\ \Eprint
  {http://arxiv.org/abs/1806.05991} {arXiv:1806.05991 [astro-ph.CO]}
  \BibitemShut {NoStop}%
\bibitem [{\citenamefont {Baum}\ \emph {et~al.}(2023)\citenamefont {Baum} \emph
  {et~al.}}]{Baum:2023cct}%
  \BibitemOpen
  \bibfield  {author} {\bibinfo {author} {\bibfnamefont {Sebastian}\
  \bibnamefont {Baum}} \emph {et~al.},\ }\bibfield  {title} {\enquote {\bibinfo
  {title} {{Mineral detection of neutrinos and dark matter. A whitepaper}},}\
  }\href {\doibase 10.1016/j.dark.2023.101245} {\bibfield  {journal} {\bibinfo
  {journal} {Phys. Dark Univ.}\ }\textbf {\bibinfo {volume} {41}},\ \bibinfo
  {pages} {101245} (\bibinfo {year} {2023})},\ \Eprint
  {http://arxiv.org/abs/2301.07118} {arXiv:2301.07118 [astro-ph.IM]}
  \BibitemShut {NoStop}%
\bibitem [{\citenamefont {Aalbers}\ \emph {et~al.}(2016)\citenamefont {Aalbers}
  \emph {et~al.}}]{DARWIN:2016hyl}%
  \BibitemOpen
  \bibfield  {author} {\bibinfo {author} {\bibfnamefont {J.}~\bibnamefont
  {Aalbers}} \emph {et~al.} (\bibinfo {collaboration} {DARWIN}),\ }\bibfield
  {title} {\enquote {\bibinfo {title} {{DARWIN: towards the ultimate dark
  matter detector}},}\ }\href {\doibase 10.1088/1475-7516/2016/11/017}
  {\bibfield  {journal} {\bibinfo  {journal} {JCAP}\ }\textbf {\bibinfo
  {volume} {11}},\ \bibinfo {pages} {017} (\bibinfo {year} {2016})}\BibitemShut
  {NoStop}%
\bibitem [{\citenamefont {Aalbers}\ \emph
  {et~al.}(2024{\natexlab{a}})\citenamefont {Aalbers} \emph
  {et~al.}}]{XLZD:2024}%
  \BibitemOpen
  \bibfield  {author} {\bibinfo {author} {\bibfnamefont {J.}~\bibnamefont
  {Aalbers}} \emph {et~al.} (\bibinfo {collaboration} {XLZD}),\ }\bibfield
  {title} {\enquote {\bibinfo {title} {{The XLZD Design Book: Towards the
  Next-Generation Dark Matter Observatory}},}\ }\href@noop {} {\  (\bibinfo
  {year} {2024}{\natexlab{a}})}\BibitemShut {NoStop}%
\bibitem [{\citenamefont {Jordan}\ \emph {et~al.}(2020)\citenamefont {Jordan},
  \citenamefont {Baum}, \citenamefont {Stengel}, \citenamefont {Ferrari},
  \citenamefont {Morone}, \citenamefont {Sala},\ and\ \citenamefont
  {Spitz}}]{Jordan:2020gxx}%
  \BibitemOpen
  \bibfield  {author} {\bibinfo {author} {\bibfnamefont {Johnathon~R.}\
  \bibnamefont {Jordan}}, \bibinfo {author} {\bibfnamefont {Sebastian}\
  \bibnamefont {Baum}}, \bibinfo {author} {\bibfnamefont {Patrick}\
  \bibnamefont {Stengel}}, \bibinfo {author} {\bibfnamefont {Alfredo}\
  \bibnamefont {Ferrari}}, \bibinfo {author} {\bibfnamefont {Maria~Cristina}\
  \bibnamefont {Morone}}, \bibinfo {author} {\bibfnamefont {Paola}\
  \bibnamefont {Sala}}, \ and\ \bibinfo {author} {\bibfnamefont {Joshua}\
  \bibnamefont {Spitz}},\ }\bibfield  {title} {\enquote {\bibinfo {title}
  {{Measuring Changes in the Atmospheric Neutrino Rate Over Gigayear
  Timescales}},}\ }\href {\doibase 10.1103/PhysRevLett.125.231802} {\bibfield
  {journal} {\bibinfo  {journal} {Phys. Rev. Lett.}\ }\textbf {\bibinfo
  {volume} {125}},\ \bibinfo {pages} {231802} (\bibinfo {year} {2020})},\
  \Eprint {http://arxiv.org/abs/2004.08394} {arXiv:2004.08394 [hep-ph]}
  \BibitemShut {NoStop}%
\bibitem [{\citenamefont {Tapia-Arellano}\ and\ \citenamefont
  {Horiuchi}(2021)}]{Tapia-Arellano:2021cml}%
  \BibitemOpen
  \bibfield  {author} {\bibinfo {author} {\bibfnamefont {Natalia}\ \bibnamefont
  {Tapia-Arellano}}\ and\ \bibinfo {author} {\bibfnamefont {Shunsaku}\
  \bibnamefont {Horiuchi}},\ }\bibfield  {title} {\enquote {\bibinfo {title}
  {{Measuring solar neutrinos over gigayear timescales with paleo
  detectors}},}\ }\href {\doibase 10.1103/PhysRevD.103.123016} {\bibfield
  {journal} {\bibinfo  {journal} {Phys. Rev. D}\ }\textbf {\bibinfo {volume}
  {103}},\ \bibinfo {pages} {123016} (\bibinfo {year} {2021})},\ \Eprint
  {http://arxiv.org/abs/2102.01755} {arXiv:2102.01755 [hep-ph]} \BibitemShut
  {NoStop}%
\bibitem [{\citenamefont {Baum}\ \emph {et~al.}(2022)\citenamefont {Baum},
  \citenamefont {Capozzi},\ and\ \citenamefont {Horiuchi}}]{Baum:2022wfc}%
  \BibitemOpen
  \bibfield  {author} {\bibinfo {author} {\bibfnamefont {Sebastian}\
  \bibnamefont {Baum}}, \bibinfo {author} {\bibfnamefont {Francesco}\
  \bibnamefont {Capozzi}}, \ and\ \bibinfo {author} {\bibfnamefont {Shunsaku}\
  \bibnamefont {Horiuchi}},\ }\bibfield  {title} {\enquote {\bibinfo {title}
  {{Rocks, water, and noble liquids: Unfolding the flavor contents of supernova
  neutrinos}},}\ }\href {\doibase 10.1103/PhysRevD.106.123008} {\bibfield
  {journal} {\bibinfo  {journal} {Phys. Rev. D}\ }\textbf {\bibinfo {volume}
  {106}},\ \bibinfo {pages} {123008} (\bibinfo {year} {2022})},\ \Eprint
  {http://arxiv.org/abs/2203.12696} {arXiv:2203.12696 [hep-ph]} \BibitemShut
  {NoStop}%
\bibitem [{\citenamefont {Caccianiga}\ \emph {et~al.}(2024)\citenamefont
  {Caccianiga}, \citenamefont {Apollonio}, \citenamefont {Mariani},
  \citenamefont {Magnani}, \citenamefont {Galelli},\ and\ \citenamefont
  {Veutro}}]{Caccianiga:2024otm}%
  \BibitemOpen
  \bibfield  {author} {\bibinfo {author} {\bibfnamefont {Lorenzo}\ \bibnamefont
  {Caccianiga}}, \bibinfo {author} {\bibfnamefont {Lorenzo}\ \bibnamefont
  {Apollonio}}, \bibinfo {author} {\bibfnamefont {Federico~Maria}\ \bibnamefont
  {Mariani}}, \bibinfo {author} {\bibfnamefont {Paolo}\ \bibnamefont
  {Magnani}}, \bibinfo {author} {\bibfnamefont {Claudio}\ \bibnamefont
  {Galelli}}, \ and\ \bibinfo {author} {\bibfnamefont {Alessandro}\
  \bibnamefont {Veutro}},\ }\bibfield  {title} {\enquote {\bibinfo {title}
  {{Sedimentary rocks from Mediterranean drought in the Messinian age as a
  probe of the past cosmic ray flux}},}\ }\href {\doibase
  10.1103/PhysRevD.110.L121301} {\bibfield  {journal} {\bibinfo  {journal}
  {Phys. Rev. D}\ }\textbf {\bibinfo {volume} {110}},\ \bibinfo {pages}
  {L121301} (\bibinfo {year} {2024})},\ \Eprint
  {http://arxiv.org/abs/2405.04908} {arXiv:2405.04908 [astro-ph.HE]}
  \BibitemShut {NoStop}%
\bibitem [{\citenamefont {Galelli}\ \emph {et~al.}(2026)\citenamefont
  {Galelli}, \citenamefont {Caccianiga}, \citenamefont {Apollonio},
  \citenamefont {Magnani},\ and\ \citenamefont {Breton}}]{Galelli:2025gss}%
  \BibitemOpen
  \bibfield  {author} {\bibinfo {author} {\bibfnamefont {Claudio}\ \bibnamefont
  {Galelli}}, \bibinfo {author} {\bibfnamefont {Lorenzo}\ \bibnamefont
  {Caccianiga}}, \bibinfo {author} {\bibfnamefont {Lorenzo}\ \bibnamefont
  {Apollonio}}, \bibinfo {author} {\bibfnamefont {Paolo}\ \bibnamefont
  {Magnani}}, \ and\ \bibinfo {author} {\bibfnamefont {Vincent}\ \bibnamefont
  {Breton}},\ }\bibfield  {title} {\enquote {\bibinfo {title} {{A volcanic
  chronosequence as a time-resolved paleo-detector array to study the
  cosmic-ray flux in the late Pleistocene and Holocene}},}\ }\href {\doibase
  10.1088/1475-7516/2026/04/023} {\bibfield  {journal} {\bibinfo  {journal}
  {JCAP}\ }\textbf {\bibinfo {volume} {04}},\ \bibinfo {pages} {023} (\bibinfo
  {year} {2026})},\ \Eprint {http://arxiv.org/abs/2510.23126} {arXiv:2510.23126
  [astro-ph.HE]} \BibitemShut {NoStop}%
\bibitem [{\citenamefont {Baum}\ \emph {et~al.}(2021)\citenamefont {Baum},
  \citenamefont {Edwards}, \citenamefont {Freese},\ and\ \citenamefont
  {Stengel}}]{Baum:2021jak}%
  \BibitemOpen
  \bibfield  {author} {\bibinfo {author} {\bibfnamefont {S.}~\bibnamefont
  {Baum}}, \bibinfo {author} {\bibfnamefont {T.~D.~P.}\ \bibnamefont
  {Edwards}}, \bibinfo {author} {\bibfnamefont {K.}~\bibnamefont {Freese}}, \
  and\ \bibinfo {author} {\bibfnamefont {P.}~\bibnamefont {Stengel}},\
  }\bibfield  {title} {\enquote {\bibinfo {title} {{New Projections for Dark
  Matter Searches with Paleo-Detectors}},}\ }\href {\doibase
  10.3390/instruments5020021} {\bibfield  {journal} {\bibinfo  {journal}
  {Instruments}\ }\textbf {\bibinfo {volume} {5}},\ \bibinfo {pages} {21}
  (\bibinfo {year} {2021})},\ \Eprint {http://arxiv.org/abs/2106.06559}
  {arXiv:2106.06559 [astro-ph.CO]} \BibitemShut {NoStop}%
\bibitem [{\citenamefont {Fung}\ \emph {et~al.}(2025)\citenamefont {Fung},
  \citenamefont {Lucas}, \citenamefont {Balogh}, \citenamefont {Leybourne},\
  and\ \citenamefont {Vincent}}]{Fung:2025cub}%
  \BibitemOpen
  \bibfield  {author} {\bibinfo {author} {\bibfnamefont {A.}~\bibnamefont
  {Fung}}, \bibinfo {author} {\bibfnamefont {T.}~\bibnamefont {Lucas}},
  \bibinfo {author} {\bibfnamefont {L.}~\bibnamefont {Balogh}}, \bibinfo
  {author} {\bibfnamefont {M.}~\bibnamefont {Leybourne}}, \ and\ \bibinfo
  {author} {\bibfnamefont {A.~C.}\ \bibnamefont {Vincent}},\ }\bibfield
  {title} {\enquote {\bibinfo {title} {{Refining the sensitivity of new physics
  searches with ancient minerals}},}\ }\href@noop {} {\bibfield  {journal}
  {\bibinfo  {journal} {Phys. Rev. D}\ }\textbf {\bibinfo {volume} {112}},\
  \bibinfo {pages} {043040} (\bibinfo {year} {2025})},\ \Eprint
  {http://arxiv.org/abs/2504.08885} {arXiv:2504.08885 [hep-ph]} \BibitemShut
  {NoStop}%
\bibitem [{\citenamefont {Cogswell}\ \emph {et~al.}(2021)\citenamefont
  {Cogswell}, \citenamefont {Goel},\ and\ \citenamefont
  {Huber}}]{Cogswell:2021qlq}%
  \BibitemOpen
  \bibfield  {author} {\bibinfo {author} {\bibfnamefont {B.~K.}\ \bibnamefont
  {Cogswell}}, \bibinfo {author} {\bibfnamefont {A.}~\bibnamefont {Goel}}, \
  and\ \bibinfo {author} {\bibfnamefont {P.}~\bibnamefont {Huber}},\ }\bibfield
   {title} {\enquote {\bibinfo {title} {{Passive Low-Energy Nuclear-Recoil
  Detection with Color Centers}},}\ }\href {\doibase
  10.1103/PhysRevApplied.16.064060} {\bibfield  {journal} {\bibinfo  {journal}
  {Phys. Rev. Applied}\ }\textbf {\bibinfo {volume} {16}},\ \bibinfo {pages}
  {064060} (\bibinfo {year} {2021})},\ \Eprint
  {http://arxiv.org/abs/2104.13926} {arXiv:2104.13926 [physics.ins-det]}
  \BibitemShut {NoStop}%
\bibitem [{\citenamefont {Araujo}\ \emph {et~al.}(2025)\citenamefont {Araujo}
  \emph {et~al.}}]{Araujo:2025lif}%
  \BibitemOpen
  \bibfield  {author} {\bibinfo {author} {\bibfnamefont {G.~R.}\ \bibnamefont
  {Araujo}} \emph {et~al.} (\bibinfo {collaboration} {PALEOCCENE}),\ }\bibfield
   {title} {\enquote {\bibinfo {title} {{Nuclear recoil detection with color
  centers in bulk lithium fluoride}},}\ }\href@noop {} {\  (\bibinfo {year}
  {2025})},\ \Eprint {http://arxiv.org/abs/2503.20732} {arXiv:2503.20732
  [physics.ins-det]} \BibitemShut {NoStop}%
\bibitem [{\citenamefont {Vladimirov}\ \emph {et~al.}(2024)\citenamefont
  {Vladimirov} \emph {et~al.}}]{Vladimirov:2024}%
  \BibitemOpen
  \bibfield  {author} {\bibinfo {author} {\bibfnamefont {N.}~\bibnamefont
  {Vladimirov}} \emph {et~al.},\ }\bibfield  {title} {\enquote {\bibinfo
  {title} {{Benchtop mesoSPIM: a next-generation open-source light-sheet
  microscope for cleared samples}},}\ }\href {\doibase
  10.1038/s41467-024-46770-2} {\bibfield  {journal} {\bibinfo  {journal}
  {Nature Communications}\ } (\bibinfo {year} {2024}),\
  10.1038/s41467-024-46770-2}\BibitemShut {NoStop}%
\bibitem [{\citenamefont {Chatrchyan}\ \emph {et~al.}(2008)\citenamefont
  {Chatrchyan} \emph {et~al.}}]{CMS:2008}%
  \BibitemOpen
  \bibfield  {author} {\bibinfo {author} {\bibfnamefont {S.}~\bibnamefont
  {Chatrchyan}} \emph {et~al.} (\bibinfo {collaboration} {CMS}),\ }\bibfield
  {title} {\enquote {\bibinfo {title} {{The CMS experiment at the CERN LHC}},}\
  }\href {\doibase 10.1088/1748-0221/3/08/S08004} {\bibfield  {journal}
  {\bibinfo  {journal} {JINST}\ }\textbf {\bibinfo {volume} {3}},\ \bibinfo
  {pages} {S08004} (\bibinfo {year} {2008})}\BibitemShut {NoStop}%
\bibitem [{\citenamefont {Aad}\ \emph {et~al.}(2008)\citenamefont {Aad} \emph
  {et~al.}}]{ATLAS:2008}%
  \BibitemOpen
  \bibfield  {author} {\bibinfo {author} {\bibfnamefont {G.}~\bibnamefont
  {Aad}} \emph {et~al.} (\bibinfo {collaboration} {ATLAS}),\ }\bibfield
  {title} {\enquote {\bibinfo {title} {{The ATLAS Experiment at the CERN Large
  Hadron Collider}},}\ }\href {\doibase 10.1088/1748-0221/3/08/S08003}
  {\bibfield  {journal} {\bibinfo  {journal} {JINST}\ }\textbf {\bibinfo
  {volume} {3}},\ \bibinfo {pages} {S08003} (\bibinfo {year}
  {2008})}\BibitemShut {NoStop}%
\bibitem [{\citenamefont {Ziegler}\ \emph {et~al.}(2010)\citenamefont
  {Ziegler}, \citenamefont {Ziegler},\ and\ \citenamefont {Biersack}}]{TRIM}%
  \BibitemOpen
  \bibfield  {author} {\bibinfo {author} {\bibfnamefont {J.~F.}\ \bibnamefont
  {Ziegler}}, \bibinfo {author} {\bibfnamefont {M.~D.}\ \bibnamefont
  {Ziegler}}, \ and\ \bibinfo {author} {\bibfnamefont {J.~P.}\ \bibnamefont
  {Biersack}},\ }\bibfield  {title} {\enquote {\bibinfo {title} {{SRIM -- The
  Stopping and Range of Ions in Matter (2010)}},}\ }\href {\doibase
  10.1016/j.nimb.2010.02.091} {\bibfield  {journal} {\bibinfo  {journal} {Nucl.
  Instrum. Meth. B}\ }\textbf {\bibinfo {volume} {268}},\ \bibinfo {pages}
  {1818--1823} (\bibinfo {year} {2010})}\BibitemShut {NoStop}%
\bibitem [{\citenamefont {Agarwal}\ \emph {et~al.}(2021)\citenamefont
  {Agarwal}, \citenamefont {Lin}, \citenamefont {Li}, \citenamefont {Stoller},\
  and\ \citenamefont {Zinkle}}]{Agarwal:2021}%
  \BibitemOpen
  \bibfield  {author} {\bibinfo {author} {\bibfnamefont {S.}~\bibnamefont
  {Agarwal}}, \bibinfo {author} {\bibfnamefont {Y.}~\bibnamefont {Lin}},
  \bibinfo {author} {\bibfnamefont {C.}~\bibnamefont {Li}}, \bibinfo {author}
  {\bibfnamefont {R.~E.}\ \bibnamefont {Stoller}}, \ and\ \bibinfo {author}
  {\bibfnamefont {S.~J.}\ \bibnamefont {Zinkle}},\ }\bibfield  {title}
  {\enquote {\bibinfo {title} {{On the use of SRIM for computing radiation
  damage exposure}},}\ }\href@noop {} {\bibfield  {journal} {\bibinfo
  {journal} {Nucl. Instrum. Meth. B}\ }\textbf {\bibinfo {volume} {503}},\
  \bibinfo {pages} {11--29} (\bibinfo {year} {2021})}\BibitemShut {NoStop}%
\bibitem [{\citenamefont {Billard}\ \emph {et~al.}(2014)\citenamefont
  {Billard}, \citenamefont {Strigari},\ and\ \citenamefont
  {Figueroa-Feliciano}}]{Billard:2013qya}%
  \BibitemOpen
  \bibfield  {author} {\bibinfo {author} {\bibfnamefont {J.}~\bibnamefont
  {Billard}}, \bibinfo {author} {\bibfnamefont {L.}~\bibnamefont {Strigari}}, \
  and\ \bibinfo {author} {\bibfnamefont {E.}~\bibnamefont
  {Figueroa-Feliciano}},\ }\bibfield  {title} {\enquote {\bibinfo {title}
  {{Implication of neutrino backgrounds on the reach of next generation dark
  matter direct detection experiments}},}\ }\href {\doibase
  10.1103/PhysRevD.89.023524} {\bibfield  {journal} {\bibinfo  {journal} {Phys.
  Rev. D}\ }\textbf {\bibinfo {volume} {89}},\ \bibinfo {pages} {023524}
  (\bibinfo {year} {2014})},\ \Eprint {http://arxiv.org/abs/1307.5458}
  {arXiv:1307.5458 [hep-ph]} \BibitemShut {NoStop}%
\bibitem [{\citenamefont {O'Hare}(2021)}]{OHare:2021utq}%
  \BibitemOpen
  \bibfield  {author} {\bibinfo {author} {\bibfnamefont {Ciaran A.~J.}\
  \bibnamefont {O'Hare}},\ }\bibfield  {title} {\enquote {\bibinfo {title}
  {{New definition of the neutrino floor for direct dark matter searches}},}\
  }\href {\doibase 10.1103/PhysRevLett.127.251802} {\bibfield  {journal}
  {\bibinfo  {journal} {Phys. Rev. Lett.}\ }\textbf {\bibinfo {volume} {127}},\
  \bibinfo {pages} {251802} (\bibinfo {year} {2021})},\ \Eprint
  {http://arxiv.org/abs/2109.03116} {arXiv:2109.03116 [hep-ph]} \BibitemShut
  {NoStop}%
\bibitem [{\citenamefont {Wang}\ \emph {et~al.}(1999)\citenamefont {Wang},
  \citenamefont {Gong}, \citenamefont {Wang},\ and\ \citenamefont
  {Ewing}}]{Wang:1999}%
  \BibitemOpen
  \bibfield  {author} {\bibinfo {author} {\bibfnamefont {L}~\bibnamefont
  {Wang}}, \bibinfo {author} {\bibfnamefont {W}~\bibnamefont {Gong}}, \bibinfo
  {author} {\bibfnamefont {S}~\bibnamefont {Wang}}, \ and\ \bibinfo {author}
  {\bibfnamefont {R~C}\ \bibnamefont {Ewing}},\ }\bibfield  {title} {\enquote
  {\bibinfo {title} {Comparison of ion-beam irradiation effects in x{sub
  2}yo{sub 4} compounds},}\ }\href {https://www.osti.gov/biblio/20005960}
  {\bibfield  {journal} {\bibinfo  {journal} {Journal of the American Ceramic
  Society}\ }\textbf {\bibinfo {volume} {82}} (\bibinfo {year}
  {1999})}\BibitemShut {NoStop}%
\bibitem [{\citenamefont {May}\ \emph {et~al.}(2000)\citenamefont {May},
  \citenamefont {Pineau~des Forêts}, \citenamefont {Flower}, \citenamefont
  {Field}, \citenamefont {Allan},\ and\ \citenamefont {Purton}}]{May:2000}%
  \BibitemOpen
  \bibfield  {author} {\bibinfo {author} {\bibfnamefont {P.~W.}\ \bibnamefont
  {May}}, \bibinfo {author} {\bibfnamefont {G.}~\bibnamefont {Pineau~des
  Forêts}}, \bibinfo {author} {\bibfnamefont {D.~R.}\ \bibnamefont {Flower}},
  \bibinfo {author} {\bibfnamefont {D.}~\bibnamefont {Field}}, \bibinfo
  {author} {\bibfnamefont {N.~L.}\ \bibnamefont {Allan}}, \ and\ \bibinfo
  {author} {\bibfnamefont {J.~A.}\ \bibnamefont {Purton}},\ }\bibfield  {title}
  {\enquote {\bibinfo {title} {Sputtering of grains in c-type shocks},}\ }\href
  {\doibase 10.1046/j.1365-8711.2000.03796.x} {\bibfield  {journal} {\bibinfo
  {journal} {Monthly Notices of the Royal Astronomical Society}\ }\textbf
  {\bibinfo {volume} {318}},\ \bibinfo {pages} {809--816} (\bibinfo {year}
  {2000})},\ \Eprint
  {http://arxiv.org/abs/https://academic.oup.com/mnras/article-pdf/318/3/809/3474381/318-3-809.pdf}
  {https://academic.oup.com/mnras/article-pdf/318/3/809/3474381/318-3-809.pdf}
  \BibitemShut {NoStop}%
\bibitem [{\citenamefont {{Bocchio}}(2014)}]{Bocchio:2014}%
  \BibitemOpen
  \bibfield  {author} {\bibinfo {author} {\bibfnamefont {Marco}\ \bibnamefont
  {{Bocchio}}},\ }\emph {\bibinfo {title} {{Modelling Dust Processing and
  Evolution in Extreme Environments as seen by Herschel Space Observatory}}},\
  \href@noop {} {Ph.D. thesis},\ \bibinfo  {school} {Institut d'Astrophysique
  Spatiale} (\bibinfo {year} {2014})\BibitemShut {NoStop}%
\bibitem [{\citenamefont {{Webmineral Mineralogy
  Database}}()}]{webmineral_olivine}%
  \BibitemOpen
  \bibfield  {author} {\bibinfo {author} {\bibnamefont {{Webmineral Mineralogy
  Database}}},\ }\href@noop {} {\enquote {\bibinfo {title} {{Olivine Mineral
  Data}},}\ }\bibinfo {howpublished}
  {\url{https://webmineral.com/data/Olivine.shtml}},\ \bibinfo {note}
  {accessed: 2026-05-07}\BibitemShut {NoStop}%
\bibitem [{\citenamefont {Nahum}\ and\ \citenamefont
  {Wiegand}(1967)}]{nahum1967fcenters}%
  \BibitemOpen
  \bibfield  {author} {\bibinfo {author} {\bibfnamefont {J.}~\bibnamefont
  {Nahum}}\ and\ \bibinfo {author} {\bibfnamefont {D.~A.}\ \bibnamefont
  {Wiegand}},\ }\bibfield  {title} {\enquote {\bibinfo {title} {Optical
  properties of some {F}-aggregate centers in {LiF}},}\ }\href {\doibase
  10.1103/PhysRev.154.817} {\bibfield  {journal} {\bibinfo  {journal} {Physical
  Review}\ }\textbf {\bibinfo {volume} {154}},\ \bibinfo {pages} {817--830}
  (\bibinfo {year} {1967})}\BibitemShut {NoStop}%
\bibitem [{\citenamefont {Baldacchini}\ \emph {et~al.}(1996)\citenamefont
  {Baldacchini}, \citenamefont {Cremona}, \citenamefont {d'Auria},
  \citenamefont {Montereali},\ and\ \citenamefont
  {Kalinov}}]{baldacchini1996f2f3}%
  \BibitemOpen
  \bibfield  {author} {\bibinfo {author} {\bibfnamefont {G.}~\bibnamefont
  {Baldacchini}}, \bibinfo {author} {\bibfnamefont {M.}~\bibnamefont
  {Cremona}}, \bibinfo {author} {\bibfnamefont {G.}~\bibnamefont {d'Auria}},
  \bibinfo {author} {\bibfnamefont {R.~M.}\ \bibnamefont {Montereali}}, \ and\
  \bibinfo {author} {\bibfnamefont {V.}~\bibnamefont {Kalinov}},\ }\bibfield
  {title} {\enquote {\bibinfo {title} {Optical bands of {F$_2$} and {F$_3^+$}
  color centers in {LiF}},}\ }\href {\doibase 10.1103/PhysRevB.54.17508}
  {\bibfield  {journal} {\bibinfo  {journal} {Physical Review B}\ }\textbf
  {\bibinfo {volume} {54}},\ \bibinfo {pages} {17508--17516} (\bibinfo {year}
  {1996})}\BibitemShut {NoStop}%
\bibitem [{\citenamefont {Zaitsev}(2001)}]{zaitsev2001optical}%
  \BibitemOpen
  \bibfield  {author} {\bibinfo {author} {\bibfnamefont {A.~M.}\ \bibnamefont
  {Zaitsev}},\ }\href {\doibase 10.1007/978-3-662-04548-0} {\emph {\bibinfo
  {title} {Optical Properties of Diamond: A Data Handbook}}}\ (\bibinfo
  {publisher} {Springer},\ \bibinfo {address} {Berlin, Heidelberg},\ \bibinfo
  {year} {2001})\BibitemShut {NoStop}%
\bibitem [{\citenamefont {Castelletto}\ and\ \citenamefont
  {Boretti}(2020)}]{castelletto2020silicon}%
  \BibitemOpen
  \bibfield  {author} {\bibinfo {author} {\bibfnamefont {S.}~\bibnamefont
  {Castelletto}}\ and\ \bibinfo {author} {\bibfnamefont {A.}~\bibnamefont
  {Boretti}},\ }\bibfield  {title} {\enquote {\bibinfo {title} {Silicon carbide
  color centers for quantum applications},}\ }\href {\doibase
  10.1088/2515-7647/ab77a2} {\bibfield  {journal} {\bibinfo  {journal} {Journal
  of Physics: Photonics}\ }\textbf {\bibinfo {volume} {2}},\ \bibinfo {pages}
  {022001} (\bibinfo {year} {2020})}\BibitemShut {NoStop}%
\bibitem [{\citenamefont {Helm}(1956)}]{Helm:1956}%
  \BibitemOpen
  \bibfield  {author} {\bibinfo {author} {\bibfnamefont {R.~H.}\ \bibnamefont
  {Helm}},\ }\bibfield  {title} {\enquote {\bibinfo {title} {{Inelastic and
  Elastic Scattering of 187-Mev Electrons from Selected Even-Even Nuclei}},}\
  }\href {\doibase 10.1103/PhysRev.104.1466} {\bibfield  {journal} {\bibinfo
  {journal} {Phys. Rev.}\ }\textbf {\bibinfo {volume} {104}},\ \bibinfo {pages}
  {1466--1475} (\bibinfo {year} {1956})}\BibitemShut {NoStop}%
\bibitem [{\citenamefont {Lewin}\ and\ \citenamefont
  {Smith}(1996)}]{LewinSmith:1996}%
  \BibitemOpen
  \bibfield  {author} {\bibinfo {author} {\bibfnamefont {J.~D.}\ \bibnamefont
  {Lewin}}\ and\ \bibinfo {author} {\bibfnamefont {P.~F.}\ \bibnamefont
  {Smith}},\ }\bibfield  {title} {\enquote {\bibinfo {title} {{Review of
  mathematics, numerical factors, and corrections for dark matter experiments
  based on elastic nuclear recoil}},}\ }\href {\doibase
  10.1016/S0927-6505(96)00047-3} {\bibfield  {journal} {\bibinfo  {journal}
  {Astropart. Phys.}\ }\textbf {\bibinfo {volume} {6}},\ \bibinfo {pages}
  {87--112} (\bibinfo {year} {1996})}\BibitemShut {NoStop}%
\bibitem [{\citenamefont {Smith}\ \emph {et~al.}(2007)\citenamefont {Smith}
  \emph {et~al.}}]{Smith:2007zzr}%
  \BibitemOpen
  \bibfield  {author} {\bibinfo {author} {\bibfnamefont {M.~C.}\ \bibnamefont
  {Smith}} \emph {et~al.},\ }\bibfield  {title} {\enquote {\bibinfo {title}
  {{The RAVE Survey: Constraining the Local Galactic Escape Speed}},}\ }\href
  {\doibase 10.1111/j.1365-2966.2007.11964.x} {\bibfield  {journal} {\bibinfo
  {journal} {Mon. Not. Roy. Astron. Soc.}\ }\textbf {\bibinfo {volume} {379}},\
  \bibinfo {pages} {755--772} (\bibinfo {year} {2007})},\ \Eprint
  {http://arxiv.org/abs/astro-ph/0611671} {arXiv:astro-ph/0611671} \BibitemShut
  {NoStop}%
\bibitem [{\citenamefont {Read}(2014)}]{Read:2014qva}%
  \BibitemOpen
  \bibfield  {author} {\bibinfo {author} {\bibfnamefont {J.~I.}\ \bibnamefont
  {Read}},\ }\bibfield  {title} {\enquote {\bibinfo {title} {{The Local Dark
  Matter Density}},}\ }\href {\doibase 10.1088/0954-3899/41/6/063101}
  {\bibfield  {journal} {\bibinfo  {journal} {J. Phys. G}\ }\textbf {\bibinfo
  {volume} {41}},\ \bibinfo {pages} {063101} (\bibinfo {year} {2014})},\
  \Eprint {http://arxiv.org/abs/1404.1938} {arXiv:1404.1938 [astro-ph.GA]}
  \BibitemShut {NoStop}%
\bibitem [{\citenamefont {Aalbers}\ \emph
  {et~al.}(2024{\natexlab{b}})\citenamefont {Aalbers} \emph
  {et~al.}}]{LZ:2024zvo}%
  \BibitemOpen
  \bibfield  {author} {\bibinfo {author} {\bibfnamefont {J.}~\bibnamefont
  {Aalbers}} \emph {et~al.} (\bibinfo {collaboration} {LZ}),\ }\bibfield
  {title} {\enquote {\bibinfo {title} {{Dark Matter Search Results from 4.2
  Tonne-Years of Exposure of the LUX-ZEPLIN (LZ) Experiment}},}\ }\href@noop {}
  {\  (\bibinfo {year} {2024}{\natexlab{b}})},\ \Eprint
  {http://arxiv.org/abs/2410.17036} {arXiv:2410.17036 [hep-ex]} \BibitemShut
  {NoStop}%
\bibitem [{\citenamefont {Zabi}\ \emph {et~al.}(2019)\citenamefont {Zabi} \emph
  {et~al.}}]{cms2018phase2daq}%
  \BibitemOpen
  \bibfield  {author} {\bibinfo {author} {\bibfnamefont {A.}~\bibnamefont
  {Zabi}} \emph {et~al.},\ }\bibfield  {title} {\enquote {\bibinfo {title} {The
  {CMS} data acquisition system for the {Phase-2} upgrade},}\ }\href {\doibase
  10.22323/1.343.0129} {\bibfield  {journal} {\bibinfo  {journal} {Proceedings
  of Science (Topical Workshop on Electronics for Particle Physics 2018)}\ }
  (\bibinfo {year} {2019}),\ 10.22323/1.343.0129},\ \bibinfo {note}
  {arXiv:1806.08975}\BibitemShut {NoStop}%
\bibitem [{\citenamefont {Hedges}(2026)}]{hedges_talk_2026}%
  \BibitemOpen
  \bibfield  {author} {\bibinfo {author} {\bibfnamefont {Samuel}\ \bibnamefont
  {Hedges}},\ }\href@noop {} {\enquote {\bibinfo {title} {Imaging nuclear
  recoil damage in minerals with light-sheet microscopy},}\ }\bibinfo
  {howpublished} {Talk presented at the MD$\nu$DM 2026 -- Mineral Detection of
  Neutrinos and Dark Matter workshop Karlsruhe, Germany} (\bibinfo {year}
  {2026})\BibitemShut {NoStop}%
\bibitem [{\citenamefont {Shapson-Coe}\ \emph {et~al.}(2024)\citenamefont
  {Shapson-Coe} \emph {et~al.}}]{ShapsonCoe:2024}%
  \BibitemOpen
  \bibfield  {author} {\bibinfo {author} {\bibfnamefont {A.}~\bibnamefont
  {Shapson-Coe}} \emph {et~al.},\ }\bibfield  {title} {\enquote {\bibinfo
  {title} {{A petavoxel fragment of human cerebral cortex reconstructed at
  nanoscale resolution}},}\ }\href {\doibase 10.1126/science.adk4858}
  {\bibfield  {journal} {\bibinfo  {journal} {Science}\ }\textbf {\bibinfo
  {volume} {384}},\ \bibinfo {pages} {eadk4858} (\bibinfo {year}
  {2024})}\BibitemShut {NoStop}%
\end{thebibliography}%

\end{document}